\documentclass[11pt]{article}
\usepackage[margin=1in]{geometry}
\usepackage{multirow}

\usepackage{graphicx}
\usepackage{subcaption}
\usepackage{amsmath}
\usepackage{amssymb}
\usepackage{amsthm}
\usepackage{color}
\allowdisplaybreaks[1]

\usepackage{caption}
\usepackage{subcaption}
\usepackage{algorithm}
\usepackage{algpseudocode}
\usepackage{hyperref}

\theoremstyle{plain}
\newtheorem{theorem}{Theorem}[section]
\newtheorem{lem}[theorem]{Lemma}

\newtheorem{cor}[theorem]{Corollary}

\usepackage{subcaption}
\usepackage{booktabs}
\usepackage{enumitem}

\theoremstyle{definition}

\theoremstyle{remark}

\def\0{{\bf 0}}
\def\1{{\bf 1}}

\def \bmat{\left[\begin{matrix}}
	\def \emat{\end{matrix}\right]}
\def \bvec{\left(\begin{matrix}}
	\def \evec{\end{matrix}\right)}

\def \xy1vec{\left[\begin{matrix}x\\y\\1\end{matrix}\right]}
\def \QED{\begin{flushright}\Halmos\end{flushright}\end{proof}}

\def \grad{\triangledown}

\def \gp3d{\grad p_3(d)}

\def \beq{\begin{equation}}
\def \eeq{\end{equation}}
\def \baeq{\begin{equation*}\begin{aligned}}
\def \eaeq{\end{aligned}\end{equation*}}
\newcommand{\baeql}[1]{\begin{equation}\label{#1}\begin{aligned}}
\def \eaeql{\end{aligned}\end{equation}}

\title{Enhancing Federated Learning Methods with Third-Order Optimization}
\author{Jeffrey Zhang\thanks{UC Riverside Department of Statistics. \url{jeffreyz@ucr.edu}}}
\date{}

\begin{document}

\maketitle

\begin{abstract}
Higher-order approximations provide more accurate representations of an objective function, and incorporating them into federated procedures may lead to more accurate statistical methods. We investigate the use of third-order approximations for federated statistical learning, the highest degree for which local minimization remains tractable without regularization. We study two existing federated algorithms that rely on second-order approximations and develop third-order analogues by replacing their second-order components with higher-order ones. We show theoretically that the resulting procedures provide improved approximation to the corresponding pooled-data estimators, and demonstrate empirically that these improvements translate into better finite-sample performance.
\end{abstract}

\section{Introduction}

Modern statistical analyses increasingly rely on large datasets, often assembled by combining information collected across multiple institutions. Direct pooling, however, is frequently impossible. In clinical research, patient-level data are subject to privacy regulations and institutional restrictions on inter-organizational sharing \cite{fleurence2014launching}. Similar constraints arise when data are proprietary or otherwise sensitive. These limitations have motivated the development of distributed inference and federated learning, in which participating sites retain their individual-level observations and communicate only statistical summaries needed for a joint analysis \cite{mcmahan2017communication,li2020federated,kairouz2021advances}.

The transition from centralized to distributed analysis also changes the relevant computational tradeoffs. Early work on distributed statistical learning primarily considered large-scale computing environments in which communication bandwidth was limited but coordination among machines was otherwise inexpensive \cite{zhang2012communication,duchi2014optimality,mackey2015distributed,rosenblatt2016optimality,lee2017communication}. In that setting, algorithms commonly favor inexpensive iterations and communicate low-dimensional quantities such as local estimators or gradients.

When privacy or data governance is the reason for distributing the data, however, the dominant cost may instead be the number of times the participating sites must coordinate. Electronic health record studies provide a representative example. Data held by hospitals may be accessible only through controlled computing environments, and the exchange of statistical summaries may require manual review, institutional approval, or secure offline procedures \cite{jochems2016distributed,duan2019odal,sadilek2021privacy,luo2022odach}. Communication latency can therefore be substantial even when the summaries themselves are not especially large. In such applications, an algorithm that performs more computation within a communication round may be preferable if it reduces the number of rounds required for a given level of accuracy.

Many statistical procedures rely on approximations to objective functions that are unavailable, difficult to evaluate, or difficult to optimize directly. Approximation is particularly important in federated learning, where the coordinating center does not have access to the pooled individual-level data and must instead work with statistical summaries communicated by the participating sites. More broadly, approximation is fundamental to a wide range of classical and modern statistical methods, including Newton and quasi-Newton methods~\cite{huang2019distributed,safaryan2021fednl,dinh2022done,elgabli2022fednew,li2024fedns}, one-step estimators~\cite{duan2019odal,luo2022odach,chendistributed}, and surrogate objectives~\cite{jordan2019communication,chen2026mosaic}. A common approach is to construct these approximations using derivative information, which provides a local representation of the underlying objective that can be optimized or otherwise manipulated in place of the original problem.

For decades, second-order approximations have provided the basic framework for local optimization, with Newton's method and procedures involving a single Newton step being ubiquitous. Second-order approximations lead to quadratic functions, whose behavior is well understood: when the Hessian is positive or negative definite, the resulting optimization problem has a unique global minimum or maximum, respectively. The analogous use of third-order information is considerably more recent~\cite{silina2022unregularized,cai2026globally}. At first glance, the situation changes completely for cubic functions. A cubic polynomial with a nonzero third-degree component is necessarily unbounded above and below, and may not possess a local minimum or maximum at all. Thus, simply increasing the degree of an otherwise familiar optimization procedure raises a fundamental question of whether the resulting problem has a meaningful solution. Recent work on the geometry and optimization of cubic polynomials has shown that, despite these difficulties, their local extrema can be characterized and computed in polynomial time through semidefinite programming~\cite{ahmadi2022complexity}. Since local minimization of degree-four polynomials is NP-hard~\cite{ahmadi2022complexity}, cubic polynomials represent the highest-degree polynomial models for which unconstrained local minimization remains tractable in general, without the introduction of regularizing terms. While many higher-order Newton-type algorithms incorporate third-order and even higher-order information, they typically introduce regularization to ensure well-posed optimization subproblems, thereby modifying the underlying Taylor approximation and degrading the quality of the approximation~\cite{nesterov2021implementable,bubeck2019near,gasnikov2019optimal}.

Federated learning provides a potentially ideal setting for the exploration of higher-order methods. First-order methods are ubiquitous in large-scale optimization~\cite{beck2017first}, including applications such as deep neural network training~\cite{kashyap2022survey}, because each update is relatively inexpensive. However, these methods may require many iterations to reach a desired level of accuracy~\cite{elgabli2022fednew}, and in a federated setting each iteration may require a new round of coordination among participating sites. Higher-order methods offer the opposite tradeoff: by incorporating additional derivative information, they construct more informative local approximations that can lead to substantially faster convergence, but at the cost of more expensive and memory-intensive updates~\cite{nesterov2021implementable,bubeck2019near,gasnikov2019optimal,silina2022unregularized,ahmadi2024higher,cai2026globally}. In federated settings, much of this additional computational burden can be borne locally by the participating sites, where even substantial increases in computation may have limited impact on overall study turnaround time relative to the coordination overhead associated with an additional synchronization round. Consequently, the relative value of computation and communication is fundamentally different from that of conventional centralized optimization: additional local computation may be well justified when it substantially reduces the number of required synchronization rounds.

In this work, we focus on the use of third-order information to extend statistical procedures that are traditionally based on second-order approximations. In particular, we consider two complementary approaches to federated inference. The first replaces the quadratic local models used in iterative optimization with cubic approximations, with the goal of reaching a desired level of accuracy in fewer synchronization rounds. The second incorporates higher-order information into communication-limited surrogate methods, with the goal of obtaining a more accurate approximation to pooled-data inference from a single synchronization round. These approaches use the same underlying principle---replacing a second-order approximation by a more informative third-order one---but exploit the resulting improvement in different ways. The iterative approach uses the additional information to accelerate convergence, while the single-round approach uses it to improve approximation accuracy under a fixed communication budget. To our knowledge, this represents the first use of unregularized cubic optimization as a technique in the statistical literature.

Our contributions are threefold. First, we develop federated implementations of third-order methods under both iterative and single-round communication structures. For iterative estimation, derivative information is aggregated across sites to construct third-order updates corresponding to those of the pooled objective. For single-round estimation, we extend existing quadratic surrogate procedures by incorporating third-order information directly into the distributed approximation and using fourth-order information for subsequent calibration. Second, we develop statistical theory for the single-round procedure and quantify the improvement obtained by increasing the approximation degree. The resulting cubic surrogate achieves a sharper approximation to the infeasible pooled-data estimator than its quadratic counterpart, and higher-order calibration improves this approximation further. Third, we study the communication--accuracy tradeoff empirically. We compare second- and third-order iterative methods through the number of synchronization rounds required to reach a given level of accuracy, and compare the communication-limited procedures through their distance from the pooled-data estimator. These comparisons also allow us to examine when an additional synchronization round is preferable to extracting more information from a single round.

The remainder of the paper is organized as follows. Section~\ref{sec:preliminaries} introduces the federated statistical setting and reviews the second- and third-order optimization tools and communication-limited surrogate methods used throughout the paper. Section~\ref{sec:methods} develops the proposed iterative and single-round third-order procedures. Section~\ref{sec:statistical-theory} establishes the statistical properties of the single-round estimators and compares their approximation to the pooled-data estimator. Section~\ref{sec:simulations} evaluates the proposed methods and their communication--accuracy tradeoffs in logistic and Cox regression settings. Section~\ref{sec:real-data} presents the data analysis, and the final section concludes.

\section{Preliminaries}
\label{sec:preliminaries}

\subsection{Federated statistical setting}
\label{subsec:federated-setting}

Suppose individual-level data are distributed across $K$ participating sites. Site $k$ contains observations $X_{1k},\ldots,X_{n_k k}$, with local sample size $n_k$ and total sample size $N=\sum_{k=1}^K n_k$. We consider estimation of a common parameter $\theta\in\Theta\subseteq\mathbb R^p$ through optimization of a criterion function $M_N(\theta)$. If the individual-level data could be pooled, the corresponding estimator would be
\[
\widehat\theta_N=\arg\max_{\theta\in\Theta}M_N(\theta).
\]
Throughout the paper, $\widehat\theta_N$ serves as the centralized benchmark. In a federated setting, however, the individual-level observations remain distributed across the participating sites, and the coordinating center must estimate $\theta$ using quantities that can be computed locally and communicated without directly pooling the data.

The methods considered in this paper focus on objectives that decompose across participating sites. Let $M_{n_k}^{(k)}(\theta)$ denote the local contribution available at site $k$. We assume that
\begin{equation}
\label{eq:additive-pooled-criterion}
M_N(\theta)=\sum_{k=1}^K M_{n_k}^{(k)}(\theta).
\end{equation}
This structure arises naturally when the pooled criterion is a sum of individual-level contributions and the observations are partitioned across sites. Although the individual local criteria remain unavailable to the coordinator, their contributions to the global estimation problem can therefore be represented through quantities computed separately at each site.

Federated procedures differ in both the location of the computation and the number of times the participating sites communicate. Under \emph{central-server coordination}, a server without access to individual-level data receives and aggregates statistical summaries returned by the participating sites \cite{zhang2012communication,duchi2014optimality,huang2019distributed,chen2022first,chen2024distributed}. Under \emph{leading-site coordination}, one participating site retains access to its own individual-level data and coordinates the analysis using information received from the remaining sites \cite{jordan2019communication,duan2022heterogeneity,luo2022odach}. We consider both coordination structures below.

A second distinction concerns the number of synchronization rounds. In an iterative procedure, the sites repeatedly evaluate local quantities at a common parameter value, communicate these quantities to the coordinator, and receive an updated parameter value for the next iteration. When the required quantities are additive across sites, this permits the corresponding pooled update to be reproduced exactly without sharing individual-level data. The procedure can therefore converge to the pooled-data estimator, but each iteration requires an additional synchronization round.

Communication-efficient methods instead seek to reduce the number of such rounds by communicating richer summaries of the local data. A common strategy is to construct a surrogate for the unavailable pooled criterion using local estimators and derivative information \cite{jordan2019communication,duan2022heterogeneity,luo2022odach,gu2023distributed}. Many accurate distributed procedures use two communication rounds: an initial round is used to obtain a common estimator, which is returned to the participating sites so that local derivatives can be evaluated at that common value in a second round. Single-round procedures are more restrictive. Because the sites cannot reevaluate their local criteria after a global estimate has been constructed, each site must instead summarize its criterion around an expansion point available before communication. The coordinator then combines these summaries to approximate $M_N$ without further access to the local data.

Between-site heterogeneity creates an additional difficulty for these communication-limited procedures. Participating sites may have different covariate distributions, baseline risks, or relationships involving nuisance parameters even when the parameter of scientific interest is shared. Existing approaches include model-specific methods such as ODACH for the Cox model \cite{luo2022odach}, leading-site methods based on surrogate efficient scores \cite{duan2022heterogeneity}, and central-server methods allowing site-specific nuisance parameters \cite{gu2023distributed}. For the single-round procedure studied in this paper, we adopt the heterogeneous semiparametric framework of \cite{chendistributed}, which permits the nuisance component of the statistical model to vary across sites while retaining a common parameter of interest.

The iterative and communication-limited settings therefore present different approximation problems. With repeated synchronization, local information can be reevaluated at each successive global iterate, and the principal question is how many communication rounds are required to reach a desired level of accuracy. With a single synchronization round, the local summaries are fixed before the global estimator is known, and the principal question is how accurately those summaries approximate the unavailable pooled criterion. These two settings provide the basis for the higher-order procedures developed in Section~\ref{sec:methods}.

\subsection{Cubic Models and Third-Order Optimization}
\label{subsec:cubic}

Cubic polynomial models arise naturally as third-order local approximations to smooth objective functions. We write a general cubic polynomial as
\begin{equation}
\label{eq:general-cubic}
p(x)
=
c
+
g^\top x
+
\frac{1}{2}H[x^{\otimes2}]
+
\frac{1}{6}T[x^{\otimes3}],
\end{equation}
where $g\in\mathbb R^p$, $H\in\mathbb R^{p\times p}$ is symmetric, and $T$ is a symmetric third-order tensor. The constant term $c$ does not affect the optimization and will generally be omitted.

As discussed above, optimization of \eqref{eq:general-cubic} differs fundamentally from the quadratic case. A genuine cubic polynomial is unbounded above and below, may fail to possess a local extremum, and even finding a critical point of a cubic polynomial is NP-hard in general \cite{ahmadi2022complexity}. Thus, replacing a quadratic approximation by a cubic one does not automatically produce a well-defined or computationally tractable optimization problem.

The local optimization problem nevertheless has considerably more structure than these global properties suggest. Because $p$ has degree three, its Hessian is affine in $x$,
\begin{equation}
\label{eq:cubic-affine-hessian}
\nabla^2p(x)=H+T[x].
\end{equation}
Consequently, the region
\begin{equation}
\label{eq:cubic-convex-region}
\mathcal S
=
\left\{
x:\nabla^2p(x)\succeq0
\right\}
\end{equation}
is a spectrahedron: it is defined by a linear matrix inequality and is therefore convex. Moreover, $p$ is convex when restricted to $\mathcal S$. A local minimum of $p$ must lie in this region, since its Hessian must be positive semidefinite. Thus, although the polynomial is globally nonconvex and unbounded, the portion of the problem relevant to local minimization has a convex feasible region on which the objective itself is convex. In this sense, local minimization of a cubic polynomial is a convex optimization problem in spirit.

This observation is what makes direct optimization of cubic models possible. Rather than attempting to globally minimize an unbounded polynomial or enumerate its critical points, one can search over the Hessian-positive-semidefinite region \eqref{eq:cubic-convex-region}. Recent results show that this structure can be represented through semidefinite optimization and that, whenever a local minimum exists, the associated semidefinite relaxation is exact and recovers the local solution \cite{ahmadi2022complexity,ahmadi2024higher}. The corresponding statements for local maximization follow by applying the same results to $-p$.

The geometry is especially useful for third-order local models. Let $M:\mathbb R^p\rightarrow\mathbb R$ be sufficiently smooth and let $\widetilde\theta$ be an expansion point. Writing $s=\theta-\widetilde\theta$, its third-order Taylor model, excluding the constant term, is
\begin{equation}
\label{eq:cubic-taylor-model}
T_3(s;\widetilde\theta)
=
\dot M(\widetilde\theta)^\top s
+
\frac{1}{2}
\ddot M(\widetilde\theta)[s^{\otimes2}]
+
\frac{1}{6}
\dddot M(\widetilde\theta)[s^{\otimes3}].
\end{equation}
Here $\dot M$, $\ddot M$, and $\dddot M$ denote the gradient, Hessian, and symmetric third-derivative tensor, respectively, and the bracket notation denotes the corresponding tensor contractions. The model in \eqref{eq:cubic-taylor-model} is therefore a cubic polynomial of exactly the form considered above.

Repeatedly optimizing these local cubic models gives a third-order analogue of Newton's method. Given a current iterate $\theta^{(t)}$, an appropriate local solution $s^{(t)}$ of
\[
s\longmapsto T_3(s;\theta^{(t)})
\]
is computed and the iterate is updated according to
\begin{equation}
\label{eq:third-order-newton-update}
\theta^{(t+1)}
=
\theta^{(t)}+s^{(t)}.
\end{equation}
Recent work establishes conditions under which these cubic subproblems are well defined and the resulting iteration converges cubically \cite{silina2022unregularized}. We rely on these existing results when constructing the iterative federated procedure below.

For numerical stability, we incorporate several practical modifications into the cubic optimization procedure. First, when the quadratic coefficient matrix is positive definite, we apply a Hessian-based linear transformation to improve the conditioning of the cubic subproblem. We then attempt to solve the semidefinite formulation with the first-order optimality constraints imposed explicitly. If this procedure fails or does not satisfy the prescribed numerical tolerances, we first consider a relaxed semidefinite formulation that omits these constraints~\cite{silina2022unregularized}, selecting among admissible solutions according to the cubic objective value. If neither formulation produces an admissible solution, we revert to a conventional second-order Newton step. These safeguards allow the procedure to continue when the cubic subproblem cannot be reliably solved, although the local cubic convergence guarantees apply to the unmodified third-order iterations.

\subsection{Single-round surrogate inference}
\label{subsec:single-round-preliminaries}

The single-round procedure developed in \cite{chendistributed} provides the starting point for the communication-limited method considered in this paper. Our work extending their algorithm is largely motivated by their comment: ``We do not pursue an improved surrogate function with best third-order approximation to $M_N(\theta)$, as a polynomial with degree three can be non-convex and numerically difficult to maximize.'' The cubic optimization results reviewed above make it possible to pursue precisely this alternative. Like the broader approach pursued here, their method is motivated by a setting in which communication rounds are substantially more restrictive than local computation or the amount of information transmitted within a round. The procedure therefore accepts the additional cost of computing and communicating higher-order derivative information in exchange for obtaining a more accurate estimator without further synchronization. In particular, derivative information through third order is communicated in a single round and used to construct an estimator that more closely approximates pooled-data inference than procedures based on lower-order summaries.

The method is developed for heterogeneous distributed inference in which the parameter of interest is common across sites but nuisance parameters may vary. At site $k$, let
\[
M_{n_k}^{(k)}(\theta,\eta_k)
=
\sum_{i=1}^{n_k}m(X_{ik};\theta,\eta_k)
\]
denote the local criterion, where $\theta\in\Theta\subseteq\mathbb R^p$ is the common parameter of interest and $\eta_k$ is a site-specific nuisance parameter. For fixed $\theta$, define
\[
\widehat\eta_{n_k}^{(k)}(\theta)
=
\arg\max_{\eta_k}
M_{n_k}^{(k)}(\theta,\eta_k),
\]
and let
\[
M_{n_k}^{(k)}(\theta)
=
M_{n_k}^{(k)}
\left\{
\theta,\widehat\eta_{n_k}^{(k)}(\theta)
\right\}
\]
denote the corresponding profile criterion. If individual-level data could be pooled, the estimator of interest would be
\[
\widehat\theta_N
=
\arg\max_{\theta\in\Theta}
\sum_{k=1}^K M_{n_k}^{(k)}(\theta).
\]

The procedure approximates this unavailable pooled criterion using a single round of communication. Each participating site first selects an expansion point and computes derivative summaries of its local profile criterion at that point. These summaries are communicated to the coordinator, which replaces each unavailable local criterion by a Taylor approximation and optimizes the resulting aggregate surrogate. Higher-order derivative information transmitted in the same communication round is then used to calibrate the surrogate estimator without returning to the participating sites. Thus, the procedure can be summarized in three steps:
\begin{enumerate}
    \item Select site-specific expansion points and compute local derivative summaries.
    \item Construct and optimize a distributed surrogate for the pooled criterion.
    \item Use higher-order derivative information to calibrate the resulting estimator.
\end{enumerate}

Although the participating sites transmit derivative information through third order, the distributed surrogate itself remains quadratic. The third-order information is used only after surrogate optimization, through the approximate one-step calibration.

\section{Third-Order Methods for Federated Inference}
\label{sec:methods}

We study two complementary approaches for incorporating third-order information into federated inference. The first is designed for communication-limited settings in which the participating sites communicate with the coordinating center only once. Building on the single-round surrogate framework reviewed in Section~\ref{subsec:single-round-preliminaries}, we increase the degree of the approximations used at the global aggregation and calibration stages. The second approach is iterative and applies third-order Newton optimization in a federated setting by aggregating derivatives evaluated at a common parameter value. The two procedures therefore use higher-order information for different purposes: the single-round method improves the approximation attainable under a fixed communication budget, while the iterative method uses more informative updates with the goal of reducing the number of synchronization rounds required to reach a desired level of accuracy.

\subsection{Single-round third-order surrogate inference}
\label{subsec:third-order-surrogate}

We first consider the communication-limited setting. We retain the communication architecture and site-specific expansion-point construction of \cite{chendistributed}, while increasing the degree of the approximations used at the global aggregation and calibration stages. We describe the preceding and proposed procedures concurrently by letting $r\in\{2,3\}$ denote the degree of the surrogate approximation. The case $r=2$ recovers the quadratic procedure of \cite{chendistributed}, while $r=3$ gives the proposed third-order procedure.

The procedure has three stages. First, each participating site selects an expansion point and computes the derivative summaries required by the coordinator. Second, the coordinator constructs an approximation to the pooled criterion, retaining any local criterion available directly and replacing each unavailable local criterion by its degree-$r$ Taylor approximation around the corresponding expansion point. Optimizing this criterion produces the distributed surrogate estimator. Third, the coordinator uses the additional derivative information transmitted by the sites to construct a higher-order local approximation around the surrogate estimator and obtains a calibrated estimator. All information required for the surrogate and calibration stages is transmitted in the initial communication round.

We consider both leading-site and central-server coordination. Under leading-site coordination, site $1$ has access to its own individual-level data and coordinates the analysis. Under central-server coordination, the coordinating center has no access to individual-level data from any participating site.

\subsubsection{Expansion points and local summaries}

Under central-server coordination, each site uses its local profile estimator as the expansion point,
\[
\widetilde\theta_k^C
=
\widehat\theta_{n_k}^{(k)}
=
\arg\max_{\theta\in\Theta}M_{n_k}^{(k)}(\theta).
\]
Under leading-site coordination, site $1$ first computes its local estimator $\widehat\theta_{n_1}^{(1)}$ and Hessian $\ddot M_{n_1}^{(1)}(\widehat\theta_{n_1}^{(1)})$, which are communicated to the remaining sites. Each site $k\geq2$ then computes the mini-pooled expansion point
\begin{equation}
\label{eq:quadratic-mini-pooled}
\widetilde\theta_k^L
=
\arg\max_{\theta\in\Theta}
\left\{
M_{n_k}^{(k)}(\theta)
+
\frac12
(\theta-\widehat\theta_{n_1}^{(1)})^\top
\ddot M_{n_1}^{(1)}(\widehat\theta_{n_1}^{(1)})
(\theta-\widehat\theta_{n_1}^{(1)})
\right\}.
\end{equation}
We retain these quadratic mini-pooled expansion points for both the quadratic and cubic procedures, isolating the effect of higher-order information at the aggregation and calibration stages.

For a degree-$r$ procedure, $r\in\{2,3\}$, each site communicates its expansion point and the derivatives of its local profile criterion through order $r+1$. Thus, the cubic procedure additionally requires the fourth derivative. No further communication of local information is required.

\subsubsection{Distributed surrogate criterion}

Let $D^jM$ denote the $j$th derivative of $M$, and define the nonconstant degree-$r$ Taylor approximation at site $k$ by
\begin{equation}
\label{eq:local-degree-r-taylor}
T_{k,r}(\theta;\widetilde\theta_k^A)
=
\sum_{j=1}^{r}
\frac{1}{j!}
D^jM_{n_k}^{(k)}(\widetilde\theta_k^A)
\left[
(\theta-\widetilde\theta_k^A)^{\otimes j}
\right],
\end{equation}
where $A\in\{L,C\}$ denotes the coordination scheme. The corresponding distributed surrogate criteria are
\begin{align}
M_{N,r}^L(\theta)
&=
M_{n_1}^{(1)}(\theta)
+
\sum_{k=2}^K
T_{k,r}(\theta;\widetilde\theta_k^L),
\label{eq:leading-degree-r-surrogate}\\
M_{N,r}^C(\theta)
&=
\sum_{k=1}^K
T_{k,r}(\theta;\widetilde\theta_k^C).
\label{eq:central-degree-r-surrogate}
\end{align}

The surrogate estimator $\check\theta_{N,r}^A$ is obtained by optimizing the corresponding criterion. For $r=2$, the central-server surrogate is quadratic and can be optimized by solving its linear first-order condition, while the leading-site surrogate retains the exact leading-site criterion and is optimized using a second-order procedure. For $r=3$, the central-server surrogate is cubic, and we recover its statistically relevant local maximizer through the semidefinite formulation of Section~\ref{subsec:cubic}. Under leading-site coordination, the surrogate is generally not a cubic polynomial because the leading-site criterion remains exact. We therefore apply third-order Newton iterations to $M_{N,3}^L$, initialized at $\check\theta_{N,2}^L$, with each cubic subproblem solved using the same semidefinite procedure.

\subsubsection{Higher-order calibration}

The calibration step consists of a second optimization that incorporates higher-order derivative information to refine the surrogate estimator. For each local site, the previously transmitted higher-order derivatives are used to generate updated lower-order derivatives, thereby constructing an updated surrogate criterion. More formally, defining
\[
d_{k,r}^A
=
\check\theta_{N,r}^A-\widetilde\theta_k^A,
\]
the updated derivative of order $q$, $1\leq q\leq r$, is
\begin{equation}
\label{eq:transported-derivatives-general}
\widehat D_{k,r}^{(q),A}
=
\sum_{j=0}^{r+1-q}
\frac{1}{j!}
D^{q+j}M_{n_k}^{(k)}(\widetilde\theta_k^A)
\left[
(d_{k,r}^A)^{\otimes j}
\right].
\end{equation}

These updated derivatives are then aggregated across the participating sites. Under leading-site coordination, the derivatives of the leading-site criterion are evaluated exactly, giving
\begin{equation}
\widehat D_{N,r}^{(q),L}
=
D^qM_{n_1}^{(1)}(\check\theta_{N,r}^L)
+
\sum_{k=2}^K
\widehat D_{k,r}^{(q),L}.
\end{equation}
Under central-server coordination, the corresponding derivatives are obtained by summing over all sites,
\begin{equation}
\widehat D_{N,r}^{(q),C}
=
\sum_{k=1}^K
\widehat D_{k,r}^{(q),C}.
\end{equation}

Using these aggregated derivatives, we construct the degree-$r$ calibration criterion
\begin{equation}
\label{eq:calibration-criterion-general}
\Psi_{N,r}^A(s)
=
\sum_{q=1}^{r}
\frac{1}{q!}
\widehat D_{N,r}^{(q),A}
[s^{\otimes q}],
\end{equation}
and obtain the calibrated estimator
\begin{equation}
\label{eq:calibrated-estimator-general}
\widehat\theta_{N,r}^A
=
\check\theta_{N,r}^A
+
\operatorname{locmax}_{s}
\Psi_{N,r}^A(s).
\end{equation}

For $r=2$, the calibration criterion is quadratic, and the correction reduces to a Hessian inversion. For $r=3$, the additional fourth-order information allows us to approximate the pooled third derivative, producing a cubic calibration criterion whose relevant local maximizer is obtained using the semidefinite procedure of Section~\ref{subsec:cubic}.

Both procedures preserve the single-round communication structure, with the additional derivative information and optimization required by the cubic procedure introducing no further communication between sites.

\subsection{Federated third-order Newton}
\label{subsec:federated-third-order-newton}

We extend federated Newton–Raphson, in which sites return gradients and Hessians evaluated at a common iterate and the coordinator aggregates them into a pooled Newton step \cite{wu2012glore,lu2015webdisco}, by replacing the quadratic model with a cubic one. The construction follows the same communication pattern: at each iteration, the coordinating center broadcasts the current parameter value, the participating sites evaluate the required derivatives of their local objectives, and the resulting quantities are aggregated to determine the next update. The only substantive change is that third-order derivative information is additionally communicated and used to construct the cubic local model.

Let
\[
M_N(\theta)
=
\sum_{k=1}^K
M_{n_k}^{(k)}(\theta)
\]
be the pooled criterion. Suppose the coordinator holds a common iterate $\theta^{(t)}$. At synchronization round $t$, this value is communicated to each participating site. Site $k$ evaluates
\[
\dot M_{n_k}^{(k)}
\left(
\theta^{(t)}
\right),
\qquad
\ddot M_{n_k}^{(k)}
\left(
\theta^{(t)}
\right),
\qquad
\dddot M_{n_k}^{(k)}
\left(
\theta^{(t)}
\right)
\]
and returns these quantities to the coordinator.

For the additive objectives considered in this paper, these local derivatives aggregate exactly to the corresponding derivatives of the pooled criterion:
\begin{align}
\dot M_N
\left(
\theta^{(t)}
\right)
&=
\sum_{k=1}^K
\dot M_{n_k}^{(k)}
\left(
\theta^{(t)}
\right),\\
\ddot M_N
\left(
\theta^{(t)}
\right)
&=
\sum_{k=1}^K
\ddot M_{n_k}^{(k)}
\left(
\theta^{(t)}
\right),\\
\dddot M_N
\left(
\theta^{(t)}
\right)
&=
\sum_{k=1}^K
\dddot M_{n_k}^{(k)}
\left(
\theta^{(t)}
\right).
\end{align}
Thus, in the settings considered here, the coordinator can construct exactly the same cubic Taylor model that would be obtained from the pooled criterion.

This exactness is a consequence of the additive structure of the objectives considered in this paper, rather than a requirement of the federated third-order Newton procedure itself. When the global objective is not additive, the same communication framework can still be used, with the aggregated local derivatives providing an approximation to the derivatives of the global objective rather than reproducing them exactly. The resulting procedure is therefore applicable more broadly, although the exact equivalence with the pooled optimization problem no longer holds.

Using these aggregated derivatives, the coordinator constructs the pooled cubic Taylor model
\begin{align}
\label{eq:federated-third-order-model}
\Phi_t(s)
={}&
\dot M_N
\left(
\theta^{(t)}
\right)^\top s
+
\frac{1}{2}
\ddot M_N
\left(
\theta^{(t)}
\right)
[s^{\otimes2}]
\\
&+
\frac{1}{6}
\dddot M_N
\left(
\theta^{(t)}
\right)
[s^{\otimes3}].
\nonumber
\end{align}
The third-order Newton step is the appropriate local maximizer
\begin{equation}
\label{eq:federated-third-order-step}
s^{(t)}
=
\operatorname{locmax}_{s}
\Phi_t(s),
\end{equation}
computed using the cubic optimization procedure reviewed in Section~\ref{subsec:cubic}. The coordinator then updates
\begin{equation}
\label{eq:federated-third-order-update}
\theta^{(t+1)}
=
\theta^{(t)}
+
s^{(t)}
\end{equation}
and broadcasts the updated parameter to begin the next synchronization round.

Federation introduces no approximation into these updates. Given the same initialization and exact derivative calculations, the aggregated quantities in \eqref{eq:federated-third-order-model} are identical to those obtained from the pooled objective, and the federated third-order Newton sequence is therefore identical to its centralized counterpart. Upon convergence, the procedure consequently targets the pooled-data estimator rather than a distinct distributed approximation.

For comparison, federated second-order Newton follows the same communication pattern but transmits only the gradient and Hessian. Its update is
\begin{equation}
\label{eq:federated-second-order-step}
s_{2}^{(t)}
=
-
\left[
\ddot M_N
\left(
\theta^{(t)}
\right)
\right]^{-1}
\dot M_N
\left(
\theta^{(t)}
\right),
\end{equation}
followed by $\theta^{(t+1)}=\theta^{(t)}+s_2^{(t)}$. Thus, second- and third-order federated Newton target exactly the same pooled estimator and differ only in the local model used to determine each global update.

The communication tradeoff is consequently different from that of the single-round procedure. Third-order Newton requires each site to evaluate and transmit a third-order derivative tensor at every iteration and requires the coordinator to solve a cubic subproblem rather than a linear system. In exchange, the more informative local model may reduce the number of iterations, and therefore synchronization rounds, required to attain a specified level of accuracy. The numerical experiments in Section~\ref{sec:simulations} evaluate this tradeoff by comparing the numbers of second- and third-order synchronization rounds required for convergence and for the iterative procedures to improve upon the single-round estimators.

\section{Statistical Theory}
\label{sec:statistical-theory}

We next study the theoretical properties of the two uses of third-order information introduced in Section~\ref{sec:methods}. The two procedures require different forms of analysis. The single-round procedure replaces unavailable local criteria by cubic Taylor approximations. Its analysis concerns the statistical approximation error between the resulting distributed estimator and the infeasible pooled-data estimator. For the iterative procedure, derivative aggregation reproduces the derivatives of the pooled criterion in the additive settings considered here. The resulting algorithm is therefore a federated implementation of third-order Newton, and its local convergence properties follow from existing higher-order Newton theory.

\subsection{Single-Round Approximation to Pooled-Data Inference}
\label{subsec:theory-oneshot}

We next consider the statistical properties of the single-round estimators. Our analysis builds directly on the framework of \cite{chendistributed}. We retain the regularity conditions used there, with the smoothness condition strengthened by one derivative order to accommodate the cubic surrogate and fourth-order calibration. For completeness, we restate the conditions below. Conditions C1--C3, C5, and C6 are unchanged from \cite{chendistributed}; Condition C4 is the higher-order analogue of the original smoothness condition, requiring fourth-order differentiability and derivative control for the cubic surrogate, together with an additional fifth-order smoothness requirement for calibration.

\begin{enumerate}[label=\textbf{C\arabic*.}, leftmargin=*]

\item There exists a positive constant $C_1$ such that the sample sizes
$n_k\equiv n_{K:k}$ for $k=1,\ldots,K$ satisfy
\[
\frac{\min\{n_1,\ldots,n_K\}}
{\max\{n_1,\ldots,n_K\}}
\ge C_1
\]
for all $K$.

\item The random criterion satisfies the first Bartlett identities, and the generalized least favorable direction $h^\star(\theta,\eta)$ exists. For every sequence $\delta_K\to0$,
\[
\sup_{\eta\in\mathcal H}
\sup_{f\in\mathcal F_{\delta_K;\theta,\eta}^{K}}
P_{\theta,\eta}f^2
\to0.
\]
Moreover, there exists $\delta>0$ such that
$\mathcal F_{\delta;\theta,\eta}^{K}$ possesses an envelope
$F_{K;\theta,\eta}(x)$ satisfying
\[
\sup_{K\in\mathbb Z_+,\eta\in\mathcal H}
P_{\theta,\eta}F_{K;\theta,\eta}^2
\le C_2^2,
\]
and
\[
\sup_{\eta\in\mathcal H}
P_{\theta,\eta}
F_{K;\theta,\eta}^2
\left\{
F_{K;\theta,\eta}
\ge
\sqrt{N/K}\epsilon
\right\}
\to0
\]
for every $\epsilon>0$. The bracketing entropy satisfies
\[
\int_0^1
\sup_{\eta\in\mathcal H,\delta\in(0,1]}
\sqrt{
\log N_{[]}
\left(
\epsilon\delta,
\mathcal F_{\delta;\theta,\eta},
L_2(P_{\theta,\eta})
\right)}
\,d\epsilon
<\infty.
\]

\item The $p\times p$ matrices
\[
\widetilde V_k
\equiv
\operatorname{var}
\{
\widetilde m_{\theta,\eta_k}(X;\theta,\eta_k)
\}
\]
and
\[
\widetilde I_k
\equiv
E_{\theta,\eta_k}
\ddot m_{12}(X;\theta,\eta_k)
[h^\star(\theta,\eta_k)]
-
E_{\theta,\eta_k}
\ddot m_{11}(X;\theta,\eta_k)
\]
for $k=1,\ldots,K$ are uniformly positive definite. In addition, the weighted averages of
$\{\widetilde I_1,\ldots,\widetilde I_K\}$ and
$\{\widetilde V_1,\ldots,\widetilde V_K\}$ converge, with limits
\[
\widetilde I
\equiv
\lim_{N\to\infty}
\frac1N\sum_{k=1}^K n_k\widetilde I_k,
\qquad
\widetilde V
\equiv
\lim_{N\to\infty}
\frac1N\sum_{k=1}^K n_k\widetilde V_k.
\]

\item \textit{(Modified.)}
For each $k=1,\ldots,K$, the profile function
$\theta\mapsto M_{n_k}^{(k)}(\theta)$ is four times continuously differentiable almost surely. The derivatives satisfy
\[
\sup_{k=1,\ldots,K}
\left\|
\ddot M_{n_k}^{(k)}(\theta)/n_k
+
\widetilde I_k
\right\|
=
o_{P_{\theta,\Gamma}}(1),
\]
\[
\sup_{k,\widetilde\theta}
\left\|
\dddot M_{n_k}^{(k)}(\widetilde\theta)/n_k
\right\|
=
O_{P_{\theta,\Gamma}}(1),
\]
and
\[
\sup_{k,\widetilde\theta}
\left\|
\ddddot M_{n_k}^{(k)}(\widetilde\theta)/n_k
\right\|
=
O_{P_{\theta,\Gamma}}(1).
\]

For the cubic calibrated estimator, we additionally assume that the profile functions are five times continuously differentiable almost surely and satisfy
\[
\sup_{k,\widetilde\theta}
\left\|
\frac{\overset{5}{M}_{n_k}^{(k)}(\widetilde\theta)}{n_k}
\right\|
=
O_{P_{\theta,\Gamma}}(1).
\]

\item The pooled-data estimator $\widehat\theta_N$ converges to $\theta$ in probability. The M-estimators
$\widehat\eta_{n_k}^{(k)}(\theta)$ given the parameter value $\theta$ at site $k$ satisfy
\[
\sup_{k=1,\ldots,K}
E\left[
n_k^{vr_1}
\left\|
\widehat\eta_{n_k}^{(k)}(\theta)-\eta_k
\right\|^{r_1}
\right]
=
O(1)
\]
for some $v\in(1/4,1/2]$ and $r_1\ge1$. Furthermore,
$\{\widehat\eta_{n_k}^{(k)}(\theta):k=1,\ldots,K\}$ satisfies, for some
$\beta\in(0,1]$ with $v(1+\beta)>1/2$,
\[
\sup_{k=1,\ldots,K}
P_{\theta,\eta_k}
\widetilde m_{\theta,\eta_k}
\left(
\cdot;
\theta,
\widehat\eta_{n_k}^{(k)}(\theta)
\right)
=
O_{P_{\theta,\Gamma}}
\left(
\sup_{k=1,\ldots,K}
\left\|
\widehat\eta_{n_k}^{(k)}(\theta)-\eta_k
\right\|^{1+\beta}
\right).
\]

\item The random variables
$\widetilde m_{\theta,\eta_k}(X_k;\theta,\eta_k)$,
$k=1,\ldots,K$, satisfy
\[
\sup_{k=1,\ldots,K}
E
\left[
\left\|
\widetilde m_{\theta,\eta_k}
(X_k;\theta,\eta_k)
\right\|^2
I\left\{
\left\|
\widetilde m_{\theta,\eta_k}
(X_k;\theta,\eta_k)
\right\|
\ge\sqrt N
\right\}
\right]
\to0.
\]

\end{enumerate}

Conditions C1--C3, C5, and C6 are identical to those of
\cite{chendistributed}. The only modification is Condition C4,
which strengthens the smoothness requirements to accommodate
the cubic surrogate and its fourth-order calibration.
The pooled-data estimator and the local and mini-pooled
expansion points are unchanged, allowing us to carry forward
their existing convergence results.

For $A\in\{L,C\}$, define
\[
d_N
=
\max_{A\in\{L,C\}}
\sup_{k\in\mathcal K_A}
\left\|
\widetilde\theta_k^A-\widehat\theta_N
\right\|,
\]
where $\mathcal K_L=\{2,\ldots,K\}$ and
$\mathcal K_C=\{1,\ldots,K\}$.

\begin{lem}[Inherited convergence results]
\label{lem:inherited-convergence}
Under Conditions C1--C6 and the baseline site-growth
conditions of \cite{chendistributed}, the pooled-data
estimator satisfies
\[
\widehat\theta_N-\theta
=
O_{P_{\theta,\Gamma}}(N^{-1/2}),
\]
and the local expansion points satisfy
\[
d_N
=
o_{P_{\theta,\Gamma}}(N^{-1/4}).
\]
Furthermore, the existing quadratic surrogate and calibrated
estimators satisfy, for $A\in\{L,C\}$,
\[
\check\theta_{N,2}^A-\widehat\theta_N
=
O_{P_{\theta,\Gamma}}(d_N^2),
\]
and
\[
\widehat\theta_{N,2}^A-\widehat\theta_N
=
O_{P_{\theta,\Gamma}}(d_N^3).
\]
\end{lem}

These results follow directly from the analysis of
\cite{chendistributed}. We therefore focus on the
additional approximation guarantees obtained by incorporating
higher-order information.

\begin{theorem}[Cubic surrogate approximation]
\label{thm:cubic-surrogate-rate}
Suppose Conditions C1--C3, C5, and C6 hold, together
with the fourth-order smoothness requirements of C4.
For each coordination scheme $A\in\{L,C\}$, the
degree-three surrogate criterion admits, with probability
tending to one, a statistically relevant local maximizer
$\check\theta_{N,3}^A$ in a neighborhood of the pooled
estimator. Moreover,
\[
\check\theta_{N,3}^A-\widehat\theta_N
=
O_{P_{\theta,\Gamma}}(d_N^3).
\]
\end{theorem}

\begin{theorem}[Fourth-order calibrated approximation]
\label{thm:cubic-calibrated-rate}
Under the conditions of Theorem~\ref{thm:cubic-surrogate-rate},
suppose additionally that the fifth-order smoothness
requirement in C4 holds. Then the calibrated cubic
estimator satisfies, for $A\in\{L,C\}$,
\[
\widehat\theta_{N,3}^A-\widehat\theta_N
=
O_{P_{\theta,\Gamma}}(d_N^4).
\]
\end{theorem}

The proofs follow the general structure of the quadratic
analysis in \cite{chendistributed}. The principal differences
are the additional third-derivative terms in the surrogate
first-order condition and the need to establish the existence
of a statistically relevant local maximizer of the cubic
criterion. Under C4, the additional curvature terms are
asymptotically negligible, while the Taylor remainder
improves by one order. The calibration step yields a further
improvement through the additional fourth-order information.
Complete proofs are provided in the Supplementary Material.

Combining these results with the inherited localization
rate gives the following consequences.

\begin{cor}[Approximation rates and asymptotic equivalence]
\label{cor:single-round-rates}
Under Conditions C1--C6 and the baseline site-growth
conditions of \cite{chendistributed}, the cubic surrogate
and calibrated estimators satisfy, for $A\in\{L,C\}$,
\[
\check\theta_{N,3}^A-\widehat\theta_N
=
o_{P_{\theta,\Gamma}}(N^{-3/4}),
\]
and
\[
\widehat\theta_{N,3}^A-\widehat\theta_N
=
o_{P_{\theta,\Gamma}}(N^{-1}).
\]
Consequently, both estimators are first-order asymptotically
equivalent to the pooled-data estimator:
\[
\sqrt N
\left(
\check\theta_{N,3}^A-\widehat\theta_N
\right)
=
o_{P_{\theta,\Gamma}}(1),
\]
and
\[
\sqrt N
\left(
\widehat\theta_{N,3}^A-\widehat\theta_N
\right)
=
o_{P_{\theta,\Gamma}}(1).
\]
In particular, they inherit the limiting distribution
of the pooled-data estimator established in
\cite{chendistributed}.
\end{cor}

Table~\ref{tab:rate-comparison-order} summarizes the
approximation rates of the quadratic and cubic procedures,
both in terms of the expansion-point discrepancy $d_N$
and the total sample size $N$.

\begin{table}[H]
\centering
\small
\begin{tabular}{llcc}
\toprule
Procedure & Estimator
& General rate & Rate in $N$ \\
\midrule
Quadratic & Surrogate
& $O_P(d_N^2)$ & $o_P(N^{-1/2})$ \\
& Calibrated
& $O_P(d_N^3)$ & $o_P(N^{-3/4})$ \\
\midrule
Cubic & Surrogate
& $O_P(d_N^3)$ & $o_P(N^{-3/4})$ \\
& Calibrated
& $O_P(d_N^4)$ & $o_P(N^{-1})$ \\
\bottomrule
\end{tabular}
\caption{Approximation rates relative to the pooled-data
estimator under quadratic and cubic surrogate constructions.
The rates in $N$ follow from the baseline site-growth
conditions of \cite{chendistributed}.}
\label{tab:rate-comparison-order}
\end{table}

Thus, the cubic surrogate attains the approximation order
of the previously calibrated quadratic estimator without
a separate calibration step, while fourth-order calibration
provides a further improvement.

\subsection{Third-order Newton under derivative aggregation}
\label{subsec:theory-newton}

Suppose that the pooled criterion is additive across sites,
\[
M_N(\theta)
=
\sum_{k=1}^K M_{n_k}^{(k)}(\theta).
\]
For each derivative order $j$ for which the local criteria are differentiable,
\[
\nabla^j M_N(\theta)
=
\sum_{k=1}^K
\nabla^j M_{n_k}^{(k)}(\theta).
\]
Consequently, when all sites evaluate their derivatives at a common iterate $\theta_t$, aggregation of the local gradient, Hessian, and third derivative exactly recovers the corresponding derivatives of the pooled criterion. The federated third-order Newton update is therefore identical to the update that would have been obtained from pooled individual-level data.

The local behavior of the resulting procedure follows from the theory of unregularized third-order Newton methods. In particular, consider a pooled criterion $M_N$ having a strict local maximizer $\widehat\theta_N$ and suppose that, in a neighborhood of $\widehat\theta_N$, the Hessian has the appropriate definite sign and the third derivative satisfies the required local Lipschitz condition. If the current iterate is sufficiently close to $\widehat\theta_N$, the cubic Taylor model admits the statistically relevant local maximizer and the next third-order Newton iterate is well defined. Moreover, the resulting sequence converges locally with order three; equivalently, there exists a constant $C>0$ such that
\[
\|\theta_{t+1}-\widehat\theta_N\|
\le
C\|\theta_t-\widehat\theta_N\|^3
\]
for iterates in a sufficiently small neighborhood of $\widehat\theta_N$.

Thus, in the additive setting, federating the procedure does not introduce an additional approximation error: the only distinction from centralized third-order Newton is that construction of each Taylor model requires a synchronization of the participating sites. The third-order convergence result therefore translates directly into a potential reduction in communication rounds relative to the quadratic convergence of classical Newton. This is the property examined empirically in Section~\ref{sec:simulations}.

\section{Simulation Study}
\label{sec:simulations}

We evaluate the two proposed uses of third-order information in logistic and Cox regression settings. The simulation designs follow those used in \cite{chendistributed}, allowing the proposed methods to be compared with the preceding procedures under the same data-generating settings. We consider both single-round and iterative procedures. For the single-round estimators, we report the mean $L_2$ distance to the pooled estimator, root mean squared approximation error (RMSE), mean standard deviation, and mean $L_2$ distance to the true parameter.

To distinguish the contributions of higher-order derivative information and higher-order optimization, we consider seven single-round estimators. We begin with the quadratic and cubic augmented surrogates, together with the third-order calibrated quadratic estimator. We then consider four additional estimators obtained by combining either the quadratic or cubic augmented surrogate with fourth-order calibration, using either a second-order Newton step or a third-order cubic optimization step. These comparisons allow us to isolate the effect of the surrogate order, the accuracy of the derivative approximation used for calibration, and the order of the calibration optimization. In particular, comparing second- and third-order calibration steps from the same starting surrogate allows us to assess the contribution of cubic optimization independently of the additional derivative information. Results are reported separately for leading-site and central-server coordination.

For the iterative procedures, we compare second- and third-order Newton methods through the number of synchronization rounds required for convergence and the number required to attain a pooled objective value exceeding that of the single-round estimator. For the iterative results, we report joint distributions of the numbers of second- and third-order Newton iterations. One table records iterations required for convergence to the pooled estimator, while the other records iterations required for the iterative estimator to achieve a pooled objective value exceeding that of the single-round estimator. Because derivative aggregation is exact for the additive objectives considered here, both iterative procedures target the same pooled estimator; the relevant comparison is therefore the number of synchronization rounds required to reach a given level of accuracy.

More detailed results are included in the Supplement in Section~\ref{sec: full experimental results}, including comparison against the use of a third-order mini-surrogate.

\subsection{Homogeneous logistic regression}
\label{subsec:sim-logistic}

We first consider a homogeneous logistic regression setting following the simulation design of \cite{chendistributed}. We simulate $K=50$ participating sites with $n_k=500$ observations per site and repeat the experiment over 500 Monte Carlo replications. Each observation contains six covariates, four generated from binomial distributions and two from normal distributions, with outcomes generated from a common logistic regression model across sites. Note that results are reported without an intercept term used in the logistic link function; results with an intercept term are not significantly different.

\subsubsection{Single-round estimation}

For each simulated dataset, we compute the infeasible pooled-data estimator and the seven single-round estimators described above. Results are reported under both leading-site and central-server coordination in Table~\ref{tab:logistic-oneshot}.

\begin{table}[H]
\centering
\small
\begin{tabular}{l|ccc|cccc}
\hline
Coordination & Surr. & Approx. & Opt. & Mean $L_2$ Diff. & RMSE & Mean SD & Mean $L_2$ Error \\
\hline
Pooled estimator & -- & -- & -- & -- & -- & 34.629 & 89.271 \\
\hline
\multirow{7}{*}{Leading site} & 2 & 0 & 0 & 46.426 & 48.789 & 34.877 & 98.641 \\
 & 2 & 3 & 2 & 2.847 & 3.903 & 34.681 & 89.339 \\
 & 3 & 0 & 0 & 2.804 & 3.935 & 34.695 & 89.375 \\
 & 2 & 4 & 2 & 0.882 & 1.155 & 34.619 & 89.232 \\
 & 2 & 4 & 3 & 0.194 & 0.317 & 34.625 & 89.258 \\
 & 3 & 4 & 2 & 0.205 & 0.341 & 34.625 & 89.257 \\
 & 3 & 4 & 3 & 0.201 & 0.331 & 34.625 & 89.257 \\
\hline
\multirow{7}{*}{Central server} & 2 & 0 & 0 & 96.092 & 96.460 & 33.838 & 125.186 \\
 & 2 & 3 & 2 & 7.645 & 7.818 & 34.540 & 89.181 \\
 & 3 & 0 & 0 & 5.110 & 5.305 & 34.569 & 89.142 \\
 & 2 & 4 & 2 & 3.636 & 3.716 & 34.585 & 89.161 \\
 & 2 & 4 & 3 & 0.881 & 0.959 & 34.618 & 89.230 \\
 & 3 & 4 & 2 & 0.953 & 1.032 & 34.617 & 89.226 \\
 & 3 & 4 & 3 & 0.948 & 1.026 & 34.617 & 89.227 \\
\hline
\end{tabular}
\caption{Simulation results for the single-round estimators in the homogeneous logistic regression setting. Surr., Approx., and Opt. denote the surrogate order, calibration approximation order, and optimization order, respectively, while mean $L_2$ error denotes the average Euclidean distance between the estimated and true parameter vectors. All reported statistics are multiplied by $1000$.}
\label{tab:logistic-oneshot}
\end{table}

The single-round results in Table~\ref{tab:logistic-oneshot} demonstrate substantial improvements in approximation accuracy from incorporating higher-order information under both leading-site and central-server coordination. As in~\cite{chendistributed}, the leading-site procedure generally provides a more accurate approximation to the pooled estimator. We also observe that the third-order calibrated quadratic estimator and the uncalibrated cubic surrogate perform similarly, consistent with the theoretical results in Section~\ref{sec:statistical-theory}. Incorporating fourth-order derivative information into the calibration step provides further improvements, particularly when combined with third-order optimization. Indeed, for both coordination schemes, cubic calibration substantially outperforms the corresponding second-order Newton calibration when applied to the quadratic surrogate, demonstrating that the benefits of higher-order information extend beyond improved derivative approximation to the optimization step itself.

Interestingly, the quadratic surrogate provides a slightly more accurate starting point for cubic calibration than the cubic surrogate, despite being substantially less accurate before calibration, under both the leading site and the central server regimes. Nevertheless, these differences are relatively small, and both approaches provide highly accurate approximations to the pooled estimator. Finally, the improvements in approximation accuracy have relatively little effect on overall estimation error, as the calibrated and cubic estimators exhibit performance close to that of the pooled estimator.

\subsubsection{Iterative estimation}

We next compare federated second- and third-order Newton methods using the same simulated datasets. Because the logistic log-likelihood is additive across sites, aggregation of the local derivatives exactly reproduces the corresponding derivatives of the pooled objective. The two iterative procedures therefore ultimately recover the same pooled-data estimator, and their primary distinction is the number of synchronization rounds required to reach it.

For each replicate, we record both the number of iterations required to satisfy the convergence criterion and the first iteration at which each method achieves a pooled log-likelihood exceeding that of the cubically calibrated single-round estimator. The resulting joint distributions are reported in Table~\ref{tab:logistic-newton}.

\begin{table}[H]
\centering
\scriptsize

\begin{minipage}[t]{0.48\textwidth}
\centering
\textbf{Iterations to convergence}\\[3pt]
\begin{tabular}{c|rrrr}
\toprule
& \multicolumn{4}{c}{Third-order iterations} \\
Second-order iterations & 3 & 4 & 5 & 6\\
\midrule
3 & 0 & 0 & 0 & 0\\
4 & 0 & 0 & 0 & 0\\
5 & 0 & 59 & 0 & 0 \\
6 & 0 & 413 & 28 & 0 \\
\bottomrule
\end{tabular}
\end{minipage}
\hfill
\begin{minipage}[t]{0.48\textwidth}
\centering
\textbf{Iterations to exceed one-shot log-likelihood}\\[3pt]
\begin{tabular}{c|rrrr}
\toprule
& \multicolumn{4}{c}{Third-order iterations} \\
Second-order iterations & 3 & 4 & 5 & 6\\
\midrule
3 & 0 & 0 & 0 & 0\\
4 & 0 & 0 & 0 & 0\\
5 & 0 & 500 & 0 & 0 \\
6 & 0 & 0 & 0 & 0 \\
\bottomrule
\end{tabular}
\end{minipage}

\caption{Joint distributions of second- and third-order Newton iteration counts in the homogeneous logistic regression setting. The left panel reports the numbers of iterations required for convergence, and the right panel reports the numbers required to achieve a pooled log-likelihood exceeding that of the cubically calibrated single-round estimator.}
\label{tab:logistic-newton}
\end{table}

The iterative results show that third-order Newton consistently requires fewer synchronization rounds than second-order Newton to converge to the pooled estimator. Thus, the additional computation and communication required within each third-order iteration translate into a reduction in the number of cross-site synchronizations required for pooled-level accuracy.

In the logistic setting, most replicates exceeded the one-shot likelihood on the final iteration for the third-order method. For this data-generating setting, the one-shot method is more effective when only a small number of communication rounds is feasible.

\subsection{Common-effects stratified Cox regression}
\label{subsec:sim-cox}

We next consider a common-effects stratified Cox regression setting, again following the simulation design of \cite{chendistributed}. We use the same random seeds to evaluate all methods on identical Monte Carlo replicates. We simulate $K=50$ participating sites with $n_k=200$ observations per site and repeat the experiment over 500 Monte Carlo replications. Each observation contains six correlated binary covariates. The regression coefficients are common across sites, while heterogeneity is introduced through site-specific Weibull baseline hazards. Thus, the regression parameter of interest is shared across the participating sites, while the underlying survival distributions are allowed to differ.

We follow~\cite{chendistributed} in imposing bounds during the optimization of the mini-surrogate estimators. The same bounds of $[-5,5]$ are applied to each regression coefficient, and simulation results without these bounds are provided in the Supplement in Section~\ref{sec: bounded vs unbounded}. These additional experiments demonstrate that, in finite samples, the benefits of incorporating higher-order information can depend on the stability of the mini-surrogate estimators. In particular, when certain covariates occur infrequently within individual sites, the resulting mini-surrogate estimates can become unstable, potentially amplifying errors in the higher-order approximations and degrading estimation performance.

\subsubsection{Single-round estimation}

As in the previous case, we compute the infeasible pooled-data estimator and the seven single-round estimators.  Results are reported under both leading-site and central-server coordination in Table~\ref{tab:cox-oneshot}.

\begin{table}[H]
\centering
\small
\begin{tabular}{l|ccc|cccc}
\hline
Coordination & Surr. & Approx. & Opt. & Mean $L_2$ Diff. & RMSE & Mean SD & Mean $L_2$ Error \\
\hline
Pooled estimator & -- & -- & -- & -- & -- & 104.797 & 251.667 \\
\hline
\multirow{7}{*}{Leading site} & 2 & 0 & 0 & 345.290 & 385.920 & 121.063 & 406.602 \\
 & 2 & 3 & 2 & 185.611 & 236.274 & 117.596 & 311.888 \\
 & 3 & 0 & 0 & 193.669 & 238.178 & 121.028 & 317.075 \\
 & 2 & 4 & 2 & 135.762 & 183.457 & 110.102 & 283.798 \\
 & 2 & 4 & 3 & 122.598 & 164.269 & 110.213 & 278.015 \\
 & 3 & 4 & 2 & 151.371 & 556.302 & 172.579 & 304.672 \\
 & 3 & 4 & 3 & 123.435 & 165.971 & 110.275 & 278.432 \\
\hline
\multirow{7}{*}{Central server} & 2 & 0 & 0 & 597.787 & 604.787 & 94.147 & 609.180 \\
 & 2 & 3 & 2 & 380.437 & 387.990 & 96.523 & 416.581 \\
 & 3 & 0 & 0 & 362.149 & 369.567 & 97.343 & 402.370 \\
 & 2 & 4 & 2 & 303.370 & 309.630 & 96.368 & 356.478 \\
 & 2 & 4 & 3 & 262.811 & 268.130 & 98.009 & 329.685 \\
 & 3 & 4 & 2 & 269.245 & 274.871 & 97.711 & 333.540 \\
 & 3 & 4 & 3 & 265.622 & 271.039 & 97.831 & 331.303 \\
\hline
\end{tabular}
\caption{Simulation results for the single-round estimators in the common-effects stratified Cox regression setting. Surr., Approx., and Opt. denote the surrogate order, calibration approximation order, and optimization order, respectively, while mean $L_2$ error denotes the average Euclidean distance between the estimated and true parameter vectors. All reported statistics are multiplied by $1000$.}
\label{tab:cox-oneshot}
\end{table}

The single-round results in Table~\ref{tab:cox-oneshot} exhibit a similar qualitative pattern to those in the logistic setting, with substantial improvements in approximation accuracy as higher-order information is incorporated under both leading-site and central-server coordination. The third-order calibrated quadratic estimator and the uncalibrated cubic surrogate again perform similarly, consistent with the theoretical results in Section~\ref{sec:statistical-theory}. Incorporating fourth-order derivative information provides further improvements, particularly when combined with third-order optimization. Indeed, for both coordination schemes and both starting surrogates, cubic calibration improves upon the corresponding second-order Newton calibration, demonstrating an additional benefit from higher-order optimization beyond improved derivative approximation.

As in the logistic setting, the quadratic surrogate provides a slightly more accurate starting point for cubic calibration than the cubic surrogate, despite being substantially less accurate before calibration. This pattern holds under both coordination schemes, although the differences between the resulting calibrated estimators are relatively small. The benefit of cubic optimization is particularly pronounced for the quadratic surrogate, while the fourth-order calibrated cubic surrogate with a second-order Newton step exhibits an unusually large RMSE under leading-site coordination. Nevertheless, leading-site coordination consistently provides a closer approximation to the pooled estimator than central-server coordination. Unlike the logistic setting, the improvements in approximation accuracy are also accompanied by more noticeable reductions in estimation error, although the calibrated estimators remain less accurate than the pooled estimator.

\subsubsection{Iterative estimation}

We next apply federated second- and third-order Newton methods to the same 500 simulated datasets. For the stratified Cox criterion considered here, aggregation of the local derivatives exactly reproduces the derivatives of the pooled objective. Both iterative procedures therefore target the same pooled-data estimator, and their primary distinction is the number of synchronization rounds required to reach it.

For each replicate, we record both the number of iterations required to satisfy the convergence criterion and the first iteration at which each method achieves a pooled log-likelihood exceeding that of the cubically calibrated single-round estimator. The resulting joint distributions are reported in Table~\ref{tab:cox-newton}.

\begin{table}[H]
\centering
\scriptsize

\begin{minipage}[t]{0.48\textwidth}
\centering
\textbf{Iterations to convergence}\\[3pt]
\begin{tabular}{c|rrrr}
\toprule
& \multicolumn{4}{c}{Third-order iterations} \\
Second-order iterations & 1 & 2 & 3 & 4\\
\midrule
1 & 0 & 0 & 0 & 0\\
2 & 0 & 0 & 0 & 0\\
3 & 0 & 303 & 0 & 0 \\
4 & 0 & 37 & 160 & 0 \\
\bottomrule
\end{tabular}
\end{minipage}
\hfill
\begin{minipage}[t]{0.48\textwidth}
\centering
\textbf{Iterations to exceed one-shot log-likelihood}\\[3pt]
\begin{tabular}{c|rrrr}
\toprule
& \multicolumn{4}{c}{Third-order iterations} \\
Second-order iterations & 1 & 2 & 3 & 4\\
\midrule
1 & 109 & 0 & 0 & 0\\
2 & 142 & 136 & 0 & 0\\
3 & 0 & 113 & 0 & 0 \\
4 & 0 & 0 & 0 & 0 \\
\bottomrule
\end{tabular}
\end{minipage}

\caption{Joint distributions of second- and third-order Newton iteration counts in the common-effects stratified Cox regression setting. The left panel reports the numbers of iterations required for convergence, and the right panel reports the numbers required to achieve a pooled log-likelihood exceeding that of the cubically calibrated single-round estimator.}
\label{tab:cox-newton}
\end{table}

The iterative results show that third-order Newton consistently requires fewer synchronization rounds than second-order Newton to converge to the pooled estimator. Thus, the additional computation and communication required within each third-order iteration translate into a reduction in the number of cross-site synchronizations required for pooled-level accuracy.

Furthermore, Newton's method displays competitive behavior even under a limited iteration budget in this setting. Though multiple iterations were needed to achieve convergence, around half of replicates achieved a better log-likelihood than the one-shot estimator within a single iteration for the third-order procedure. Compared to the logistic setting, the Newton method was more effective within fewer iterations in the Cox setting.

\subsection{Discussion}

Taken together, the simulation results demonstrate the two complementary uses of third-order information considered in this paper. In the single-round setting, increasing the approximation order consistently improves agreement with the pooled-data estimator, with the cubically calibrated procedure providing the most accurate approximation in both models. In the iterative setting, third-order Newton consistently requires fewer synchronization rounds than its second-order counterpart while targeting the same pooled estimator. The comparison between the iterative and single-round procedures further shows that the value of additional synchronization is setting dependent: in the logistic regression setting, several iterations are generally required before the iterative procedures improve upon the one-shot estimator, whereas in the Cox setting third-order Newton frequently becomes competitive after only one or two rounds.

These results highlight the communication--accuracy tradeoff motivating the proposed methods. When communication is restricted to a single round, higher-order information can be used to extract a more accurate approximation from the available summaries. When repeated synchronization is feasible, the same higher-order principle can instead be used to reduce the number of rounds required to approach pooled-data inference. The appropriate use of third-order information therefore depends on the relative costs of local computation, communication, and approximation error in the application at hand.

\section{Real-Data Experiments}
\label{sec:real-data}

We evaluate the proposed methods on two publicly available datasets, considering logistic regression and Cox proportional hazards regression. For each dataset, we examine two federated data-partitioning schemes. The first uses existing grouping information to construct heterogeneous participating sites, while the second randomly partitions the observations across sites while preserving their original sample sizes. This allows us to investigate the effects of site heterogeneity on the approximation accuracy of the distributed estimators. As in the simulation experiments, our primary measure of accuracy is the $L_2$ distance to the corresponding pooled-data estimator.

For logistic regression, we consider the \texttt{ijcnn1} dataset~\cite{prokhorov2001ijcnn}, obtained from the LIBSVM~\cite{chang2011libsvm} collection. We use the first ten indicator variables to partition the observations into ten sites, retaining the remaining twelve variables as regression covariates. The resulting dataset contains $N=35{,}000$ observations, with $3{,}500$ observations at each site. For Cox regression, we consider the Molecular Taxonomy of Breast Cancer International Consortium (METABRIC) dataset~\cite{curtis2012genomic}, which contains clinical and gene expression measurements from breast cancer patients. We follow the preprocessing procedure of Katzman et al.~\cite{katzman2018deepsurv}, retaining nine regression covariates. The resulting dataset contains $N=1{,}904$ observations distributed across five cohorts, with respective sample sizes of 734, 481, 286, 236, and 167. We use these cohort assignments to define the heterogeneous participating sites.

For both datasets, the largest site is designated as the leading site. We consider both the original heterogeneous partitions and ten randomized partitions preserving the original site sample sizes. For METABRIC, the pooled reference estimator is obtained by maximizing the stratified Cox partial likelihood, with risk sets defined separately within each participating site. Consequently, the pooled reference estimator is recomputed for each randomized partition. The characteristics of the two datasets are summarized in Table~\ref{tab:real_data_summary}.

\begin{table}[H]
\centering
\small
\begin{tabular}{lrrrl}
\hline
Dataset & $N$ & $d$ & $K$ & Site sample sizes \\
\hline
\texttt{ijcnn1} & 35,000 & 12 & 10 & $3{,}500 \times 10$ \\
METABRIC & 1,904 & 9 & 5 & 734, 481, 286, 236, 167 \\
\hline
\end{tabular}
\caption{Characteristics of the datasets used in the real-data experiments. Here, $N$ denotes the total sample size, $d$ the number of regression covariates, and $K$ the number of participating sites. Site sample sizes are reported in descending order.}
\label{tab:real_data_summary}
\end{table}

We compare the quadratic and cubic augmented surrogates, the quadratic surrogate with third-order calibration, and the quadratic and cubic surrogates with fourth-order calibration. The results are reported separately for leading-site and central-server coordination. For the heterogeneous partitions, we report the $L_2$ distance to the pooled estimator, while for the randomized partitions, we report the mean, median, and interquartile range across 20 random allocations.

\begin{table}[H]
\centering
\small
\begin{tabular}{l|ccc|c|cccc}
\hline
& \multicolumn{3}{c|}{} & \textbf{Heterogeneous} & \multicolumn{4}{c}{\textbf{Randomized}} \\
Coordination & Surr. & Approx. & Opt. & $L_2$ & Mean & Median & 25\% & 75\% \\
\hline
\multicolumn{9}{l}{\textbf{ijcnn1}} \\
\hline
\multirow{7}{*}{Leading site} & 2 & 0 & 0 & 636.722 & 20.101 & 20.456 & 17.002 & 22.981 \\
 & 2 & 3 & 2 & 434.499 & 1.199 & 1.220 & 0.577 & 1.657 \\
 & 3 & 0 & 0 & 411.494 & 1.204 & 1.108 & 0.625 & 1.721 \\
 & 2 & 4 & 2 & 353.538 & 0.409 & 0.378 & 0.274 & 0.483 \\
 & 2 & 4 & 3 & 267.638 & 0.225 & 0.207 & 0.132 & 0.272 \\
 & 3 & 4 & 2 & 294.620 & 0.227 & 0.209 & 0.133 & 0.277 \\
 & 3 & 4 & 3 & 281.024 & 0.226 & 0.209 & 0.133 & 0.274 \\
\hline
\multirow{7}{*}{Central server} & 2 & 0 & 0 & 937.629 & 51.921 & 52.754 & 49.182 & 57.029 \\
 & 2 & 3 & 2 & 695.012 & 2.383 & 2.208 & 1.413 & 2.879 \\
 & 3 & 0 & 0 & 678.457 & 2.095 & 1.893 & 1.314 & 2.551 \\
 & 2 & 4 & 2 & 524.007 & 2.287 & 2.240 & 1.776 & 2.683 \\
 & 2 & 4 & 3 & 411.609 & 1.121 & 1.034 & 0.731 & 1.412 \\
 & 3 & 4 & 2 & 450.041 & 1.151 & 1.072 & 0.732 & 1.451 \\
 & 3 & 4 & 3 & 420.510 & 1.149 & 1.067 & 0.732 & 1.448 \\
\hline
\multicolumn{9}{l}{\textbf{METABRIC}} \\
\hline
\multirow{7}{*}{Leading site} & 2 & 0 & 0 & 50.291 & 20.761 & 18.779 & 13.858 & 23.534 \\
 & 2 & 3 & 2 & 20.897 & 2.946 & 1.975 & 1.184 & 4.366 \\
 & 3 & 0 & 0 & 20.971 & 2.961 & 1.953 & 1.177 & 4.328 \\
 & 2 & 4 & 2 & 1.418 & 2.195 & 1.466 & 0.967 & 3.595 \\
 & 2 & 4 & 3 & 1.420 & 2.197 & 1.465 & 0.967 & 3.607 \\
 & 3 & 4 & 2 & 1.421 & 2.197 & 1.465 & 0.967 & 3.606 \\
 & 3 & 4 & 3 & 1.420 & 2.197 & 1.465 & 0.967 & 3.607 \\
\hline
\multirow{7}{*}{Central server} & 2 & 0 & 0 & 304.047 & 176.745 & 182.494 & 121.155 & 225.035 \\
 & 2 & 3 & 2 & 67.926 & 33.642 & 19.435 & 11.274 & 45.060 \\
 & 3 & 0 & 0 & 68.396 & 35.717 & 20.112 & 11.808 & 47.714 \\
 & 2 & 4 & 2 & 10.364 & 12.129 & 6.825 & 2.455 & 13.837 \\
 & 2 & 4 & 3 & 10.150 & 11.322 & 5.912 & 2.453 & 11.679 \\
 & 3 & 4 & 2 & 10.170 & 11.428 & 5.991 & 2.456 & 11.829 \\
 & 3 & 4 & 3 & 10.183 & 11.357 & 5.979 & 2.456 & 11.813 \\
\hline
\end{tabular}
\caption{Euclidean distance to the pooled estimator under heterogeneous and randomized site partitions. Surr., Approx., and Opt. denote the surrogate order, calibration approximation order, and optimization order, respectively. Randomized results report summaries across random partitions preserving the heterogeneous site sizes. Distances are multiplied by $100$ for \texttt{ijcnn1} and $10,000$ for METABRIC.}
\label{tab:real_data_accuracy}
\end{table}

The results in Table~\ref{tab:real_data_accuracy} demonstrate substantial improvements in approximation accuracy from incorporating higher-order information under both logistic and Cox regression settings. As in the simulation experiments, the third-order calibrated quadratic estimator and the uncalibrated cubic surrogate exhibit similar performance, consistent with the theoretical results in Section~\ref{sec:statistical-theory}. Incorporating fourth-order derivative information provides further improvements, particularly when combined with third-order optimization. Leading-site coordination generally provides a closer approximation to the pooled estimator than central-server coordination, consistent with the additional information available to the coordinating site.

The benefits of cubic optimization are particularly pronounced for \texttt{ijcnn1}. Under both coordination schemes and partitioning settings, cubic calibration substantially improves upon the corresponding second-order Newton calibration when applied to the quadratic surrogate, demonstrating an additional benefit from higher-order optimization beyond improved derivative approximation. Interestingly, the quadratic surrogate also provides a more accurate starting point for cubic calibration than the cubic surrogate, despite being substantially less accurate before calibration. This pattern is consistent with the logistic simulation results and suggests that greater initial approximation accuracy does not necessarily translate into greater accuracy following a higher-order calibration step. For METABRIC, the differences between the fourth-order calibrated estimators are considerably smaller, particularly under leading-site coordination, where all four estimators perform nearly identically. Under central-server coordination, cubic calibration provides a modest improvement over second-order Newton calibration when applied to the quadratic surrogate.

We also observe differences between the heterogeneous and randomized partitioning schemes. For \texttt{ijcnn1}, the heterogeneous partition produces substantially larger approximation errors than the randomized partitions, indicating that differences in the covariate distributions across sites can have a considerable effect on the accuracy of the surrogate approximations. For METABRIC, the effect of partitioning is less uniform, particularly among the fourth-order calibrated estimators. Nevertheless, the improvements associated with higher-order information persist across both datasets and partitioning schemes, although the incremental benefits of cubic optimization depend on the underlying problem.


\begin{figure}[H]
\centering
\includegraphics[width=\textwidth]{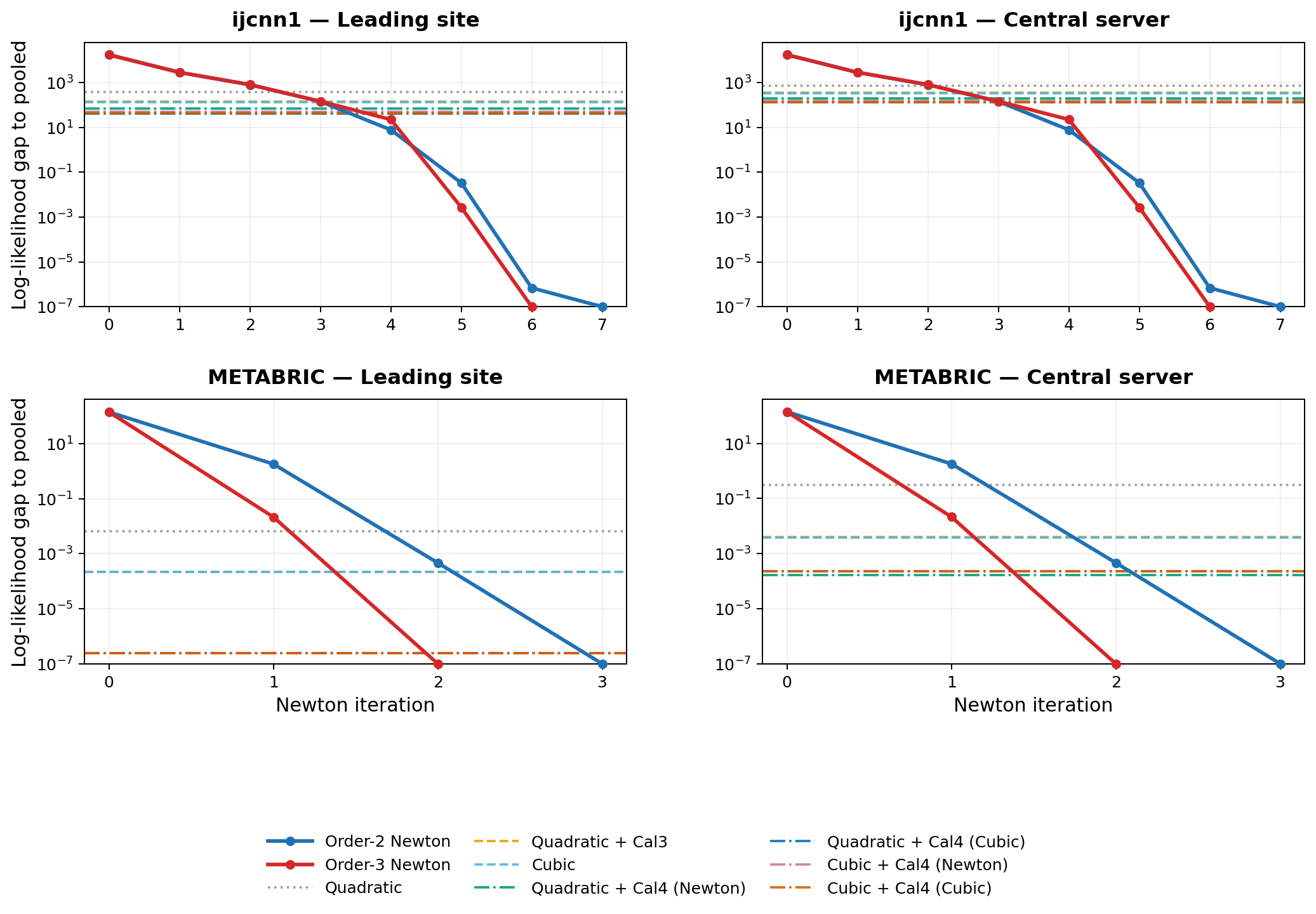}
\caption{Log-likelihood gaps to the converged pooled estimator for second- and third-order Newton methods and the seven single-round estimators under leading-site and central-server coordination. Final converged gaps are displayed at a numerical floor of $10^{-7}$.}
\label{fig:real_data_newton}
\end{figure}

We additionally compare the convergence behavior of the iterative second- and third-order Newton procedures in Figure~\ref{fig:real_data_newton}. For \texttt{ijcnn1}, third-order Newton converges in six iterations compared with seven for second-order Newton, while for METABRIC, the corresponding iteration counts are two and three. In both datasets, the iterative procedures achieve log-likelihood values exceeding those of the single-round estimators before satisfying their final convergence criteria. In particular, for \texttt{ijcnn1}, both iterative methods surpass the single-round estimators after approximately four iterations, despite requiring additional iterations for full convergence. The advantage of third-order Newton is more pronounced for METABRIC, where it achieves a substantially smaller log-likelihood gap after the first iteration. These findings demonstrate that the iterative procedures can achieve substantial improvements over single-round estimation without requiring full convergence, while third-order optimization can further reduce the number of iterations needed to attain a highly accurate solution.

\section{Conclusion}
\label{sec:conclusion}

In this work, we investigated the use of higher-order derivative information in federated estimation, focusing on both single-round surrogate methods and iterative Newton procedures. For single-round estimation, we extended existing quadratic surrogate methods by incorporating cubic approximations and fourth-order calibration, establishing improved convergence rates to the pooled-data estimator. For iterative estimation, we examined third-order Newton methods based on aggregated local derivatives, demonstrating that higher-order information can reduce the number of communication rounds required to recover the pooled estimator. Through extensive simulations and experiments on publicly available datasets, we found that incorporating higher-order information generally improves approximation accuracy in the single-round setting and communication efficiency in the iterative setting, although these improvements depend on the underlying data and computational setting.

Using this additional information is not without its disadvantages. Aside from the additional costs of optimizing the associated functions, there are also increased costs arising from computing the derivatives themselves and transmitting them. These costs become increasingly significant as the parameter dimension grows, particularly for third- and fourth-order derivative tensors. Furthermore, we observed greater potential instability in finite-sample performance, where the usage of higher-order polynomials can result in more extreme deviations from target objective functions when local estimators are highly dispersed. This concern may be particularly pronounced when local sample sizes are small, as the number of distinct entries in higher-order derivative tensors can exceed the number of observations, potentially increasing sensitivity to finite-sample variation and introducing additional risks of privacy leakage through the transmission of increasingly detailed summaries of local data.

Several directions for future research arise from these limitations. One important direction is the development of more computationally efficient methods for optimizing cubic surrogate functions, particularly approaches that avoid reliance on semidefinite programming. Exploiting the structure of higher-order derivative tensors or developing specialized interior point methods may also offer opportunities to reduce computational and communication costs. Conversely, when computational and communication costs are not primary concerns, the proposed framework can in principle be extended to incorporate derivatives of arbitrarily high order, provided sufficient smoothness conditions hold. Investigating the theoretical and practical benefits of such extensions represents another interesting direction for future research.

\section*{Acknowledgments}

We are grateful to Yong Chen, Jinyang Wang, and Yue Wu for generously sharing an at-the-time unpublished manuscript and accompanying simulation code, which provided a valuable starting point for the development and evaluation of the methods presented in this work.

\bibliographystyle{abbrv}
\bibliography{refs}

\appendix

\section{Proofs of the Single-Round Results}
\label{sec:supp-proof}

\subsection{Results from Theorem 4.1 of \cite{chendistributed}}

Throughout this section, we work under Conditions C1--C6, with Condition C4 replaced by its fourth-order analogue stated in Section~\ref{subsec:theory-oneshot}. The asymptotic properties of the infeasible pooled-data estimator are unchanged by the proposed cubic extension and have already been established in \cite{chendistributed}. We therefore do not repeat their proof. Instead, we record the consequences that will be used below.

We first recall the notation for the estimators used throughout the proof. There are $K$ participating sites, with site $k$ containing $n_k$ observations and total sample size $N=\sum_{k=1}^K n_k$. The infeasible pooled-data estimator is denoted by $\widehat\theta_N$. For each site $k$, $\widehat\theta_{n_k}^{(k)}$ denotes the local profile estimator. Under leading-site coordination, $\check\theta^{(k)}$ denotes the mini-pooled estimator used as the expansion point for $k\geq 2$. Under central-server coordination, the expansion points are instead the local profile estimators themselves. We denote the expansion points by $\widetilde\theta_k^A$, where $A\in\{L,C\}$, so that $\widetilde\theta_k^L=\check\theta^{(k)}$ for $k\geq 2$ and $\widetilde\theta_k^C=\widehat\theta_{n_k}^{(k)}$ for $k=1,\ldots,K$. Under leading-site coordination, $\check\theta_N^L$ denotes the estimator obtained by optimizing the global surrogate, while $\widehat\theta_N^L$ denotes the corresponding estimator after calibration. Analogously, under central-server coordination, $\check\theta_N^C$ and $\widehat\theta_N^C$ denote the surrogate and calibrated estimators, respectively. When it is useful to refer to either coordination scheme simultaneously, we write $\check\theta_N^A$ and $\widehat\theta_N^A$, where $A\in\{L,C\}$. 

For the proposed higher-order procedures, we append a subscript $3$ to distinguish the cubic construction from the preceding quadratic procedure. Thus, $\check\theta_{N,3}^A$ denotes the estimator obtained from the cubic surrogate and $\widehat\theta_{N,3}^A$ denotes its fourth-order calibrated counterpart. The local expansion points $\check\theta^{(k)}$ are unchanged from \cite{chendistributed}; in particular, the proposed cubic leading-site procedure continues to use the same quadratic mini-pooled estimators as its expansion points.

Throughout this section, probability statements are taken under the joint measure $P_{\theta,\Gamma}=P_{\theta,\eta_1}\times\cdots\times P_{\theta,\eta_K}$, as in \cite{chendistributed}. We work under Conditions C1--C6 stated in the main text, with C4 strengthened to its fourth-order analogue.

Under Conditions C1--C5 and the corresponding condition on $(N,K)$, Theorem 4.1 of \cite{chendistributed} gives the asymptotic expansion
\[
\widehat\theta_N-\theta
=
\left(
\sum_{k=1}^K n_k\widetilde I_k
\right)^{-1}
\sum_{k=1}^K
n_k
P_{n_k}^{(k)}
\widetilde m_{\theta,\eta_k}
(\theta,\eta_k)
+
o_{P_{\theta,\Gamma}}(N^{-1/2}),
\]
and hence
\[
\|\widehat\theta_N-\theta\|
=
O_{P_{\theta,\Gamma}}(N^{-1/2}).
\]
Under the additional Lindeberg condition C6,
\[
\sqrt{N}(\widehat\theta_N-\theta)
\rightsquigarrow
N\left(
0,
\widetilde I^{-1}\widetilde V\widetilde I^{-1}
\right).
\]
These results are unaffected by the proposed extension because the pooled-data estimator itself is unchanged.

The proof of Theorem 4.2 in~\cite{chendistributed} also provides the uniform localization results needed for the present analysis. In particular,
\[
\sup_{k=1,\ldots,K}
\left\|
\widehat\theta_{n_k}^{(k)}-\theta
\right\|
=
o_{P_{\theta,\Gamma}}(N^{-1/4}).
\]
Since the expansion points are constructed exactly as in~\cite{chendistributed}, the corresponding results for the leading-site and central-server constructions give
\[
\sup_{k=2,\ldots,K}
\left\|
\widetilde\theta_k^L-\theta
\right\|
=
o_{P_{\theta,\Gamma}}(N^{-1/4}),
\]
and
\[
\sup_{k=1,\ldots,K}
\left\|
\widetilde\theta_k^C-\theta
\right\|
=
o_{P_{\theta,\Gamma}}(N^{-1/4}).
\]
Together with $\|\widehat\theta_N-\theta\|=O_{P_{\theta,\Gamma}}(N^{-1/2})$, the triangle inequality therefore gives, for $A\in\{L,C\}$,
\[
\sup_{k\in\mathcal K_A}
\left\|
\widetilde\theta_k^A-\widehat\theta_N
\right\|
\le
\sup_{k\in\mathcal K_A}
\left\|
\widetilde\theta_k^A-\theta
\right\|
+
\left\|
\widehat\theta_N-\theta
\right\|
=
o_{P_{\theta,\Gamma}}(N^{-1/4}),
\]
where $\mathcal K_L=\{2,\ldots,K\}$ and $\mathcal K_C=\{1,\ldots,K\}$. Consequently, with probability tending to one, the pooled-data estimator and all relevant expansion points lie in a common neighborhood of the true parameter.

\subsection{Convergence Rate of the Leading-Site Cubic Surrogate Estimator}
\label{subsec:proof-cubic-consistency-leading}

Our goal is to establish the convergence rate of the leading-site cubic surrogate estimator $\check\theta_{N,3}^{L}$. Throughout the proof, we use the pooled-data asymptotic result of Theorem 4.1 and the uniform localization of the local and mini-pooled estimators established in the proof of Theorem 4.2 of~\cite{chendistributed}.

\subsubsection{Convergence Rate of the Leading-Site Surrogate Estimator}
\label{subsubsec:proof-surrogate-rate-leading}

We first establish the existence and localization of a statistically relevant local maximizer of the leading-site cubic surrogate criterion, denoted by $\check\theta_{N,3}^{L}$, before deriving its convergence rate. In~\cite{chendistributed}, the corresponding result for the second-order surrogate is obtained by first establishing the appropriate local curvature and then using an objective-value comparison to localize a maximizer of the surrogate criterion. We follow the same general proof strategy, although the presence of cubic terms requires a different treatment of both steps. We first show that the cubic perturbation does not destroy the local concavity inherited from the pooled criterion. We then show that, on the boundary of a sufficiently small neighborhood of the pooled-data estimator, the quadratic decrease in the pooled criterion dominates the higher-order approximation error of the cubic surrogate. This forces a maximizer of the surrogate criterion to lie in the interior of the neighborhood.

We first establish that the cubic surrogate retains the local concavity of the pooled criterion. Recall that its Hessian is given by
\[
\widehat H_{N,3}^L(\widetilde\theta)
=
\ddot M_{n_1}^{(1)}(\widetilde\theta)
+
\sum_{k=2}^K
\left\{
\ddot M_{n_k}^{(k)}(\check\theta^{(k)})
+
\dddot M_{n_k}^{(k)}(\check\theta^{(k)})
\left[
\widetilde\theta-\check\theta^{(k)}
\right]
\right\}.
\]
Since the leading-site criterion is retained without approximation, the difference between the cubic surrogate Hessian and the pooled Hessian is
\[
\widehat H_{N,3}^L(\widetilde\theta)
-
\ddot M_N(\widetilde\theta)
\\
=\sum_{k=2}^K
\Bigg\{
\ddot M_{n_k}^{(k)}(\check\theta^{(k)})
+
\dddot M_{n_k}^{(k)}(\check\theta^{(k)})
\left[
\widetilde\theta-\check\theta^{(k)}
\right]
-
\ddot M_{n_k}^{(k)}(\widetilde\theta)
\Bigg\}.
\]

Applying the second-order Taylor remainder to $\ddot M_{n_k}^{(k)}(\widetilde\theta)$ around $\check\theta^{(k)}$ gives
\[
\ddot M_{N,3}^L(\widetilde\theta)-\ddot M_N(\widetilde\theta)
=
-\sum_{k=2}^K
\int_0^1(1-u)
\ddddot M_{n_k}^{(k)}
\left(
\check\theta^{(k)}
+
u\left(
\widetilde\theta-\check\theta^{(k)}
\right)
\right)
\,du
\left[
\left(
\widetilde\theta-\check\theta^{(k)}
\right)^{\otimes2}
\right].
\]

Consequently, for $\mathcal B_N=B_{r_N}(\widehat\theta_N)$, we have
\[
\begin{aligned}
\sup_{\widetilde\theta\in\mathcal B_N}
\left\|
\frac{
\ddot M_{N,3}^L(\widetilde\theta)
-
\ddot M_N(\widetilde\theta)
}{N}
\right\|
\le{}&
\frac12
\sup_{k=2,\ldots,K}
\sup_{\vartheta\in\Theta}
\left\|
\frac{\ddddot M_{n_k}^{(k)}(\vartheta)}{n_k}
\right\|
\left(
r_N+
\sup_{k\ge2}
\left\|
\widehat\theta_N-\check\theta^{(k)}
\right\|
\right)^2
\\
={}&
O_{P_{\theta,\Gamma}}(1)
\left(
r_N+
\sup_{k\ge2}
\left\|
\widehat\theta_N-\check\theta^{(k)}
\right\|
\right)^2.
\end{aligned}
\]

In other words, the cubic approximation contributes only a second-order perturbation to the normalized Hessian within the neighborhood $\mathcal B_N$. We now show that this perturbation is asymptotically negligible relative to the curvature of the pooled criterion. By Conditions C3 and C4,
\[
\left\|
\frac{1}{N}\ddot M_N(\theta)+\widetilde I
\right\|
=
o_{P_{\theta,\Gamma}}(1).
\]
Moreover, the uniform third-derivative bound in C4 and the consistency of $\widehat\theta_N$ imply that, for any sequence $r_N\to0$,
\[
\begin{aligned}
\sup_{\widetilde\theta\in\mathcal B_N}
\left\|
\frac{1}{N}\ddot M_N(\widetilde\theta)+\widetilde I
\right\|
\le{}&
\left\|
\frac{1}{N}\ddot M_N(\theta)+\widetilde I
\right\|
\\
&+
O_{P_{\theta,\Gamma}}(1)
\left(
r_N+\|\widehat\theta_N-\theta\|
\right)
\\
={}&
o_{P_{\theta,\Gamma}}(1).
\end{aligned}
\]
Combining this with the preceding bound on the cubic surrogate Hessian gives
\[
\begin{aligned}
\sup_{\widetilde\theta\in\mathcal B_N}
\left\|
\frac{1}{N}\ddot M_{N,3}^L(\widetilde\theta)+\widetilde I
\right\|
\le{}&
\sup_{\widetilde\theta\in\mathcal B_N}
\left\|
\frac{\ddot M_{N,3}^L(\widetilde\theta)-\ddot M_N(\widetilde\theta)}{N}
\right\|
\\
&+
\sup_{\widetilde\theta\in\mathcal B_N}
\left\|
\frac{1}{N}\ddot M_N(\widetilde\theta)+\widetilde I
\right\|
\\
={}&
o_{P_{\theta,\Gamma}}(1).
\end{aligned}
\]
Since $\widetilde I$ is positive definite by C3, let $c_0=\lambda_{\min}(\widetilde I)>0$. By the preceding uniform convergence and the eigenvalue perturbation inequality, with probability tending to one,
\[
-\frac{1}{N}\ddot M_N(\widetilde\theta)
\succeq
\frac{c_0}{2}I,
\qquad
-\frac{1}{N}\ddot M_{N,3}^L(\widetilde\theta)
\succeq
\frac{c_0}{2}I,
\]
uniformly for $\widetilde\theta\in\mathcal B_N$. Thus, the cubic surrogate is uniformly strongly concave within this neighborhood.

Now that we have established local concavity of the cubic surrogate criterion, it remains to establish the existence of a local maximum. We do so by showing that the surrogate criterion takes strictly smaller values on the boundary of a sufficiently small neighborhood of the pooled-data estimator than at the pooled-data estimator itself. Combined with continuity of the surrogate criterion and its local concavity, this implies that the surrogate criterion admits a local maximum in the interior of the neighborhood. Let
\[
\mathcal B_N
=
\left\{
\widetilde\theta:
\left\|
\widetilde\theta-\widehat\theta_N
\right\|
\le \delta_N
\right\},
\]
and consider $\widetilde\theta\in\partial\mathcal B_N$, so that $\|\widetilde\theta-\widehat\theta_N\|=\delta_N$. By the uniform concavity established above,
\[
M_N(\widetilde\theta)
\le
M_N(\widehat\theta_N)
-
\frac{c_0N}{4}\delta_N^2.
\]

On the other hand, the cubic surrogate differs from the pooled criterion only through the Taylor approximations of the non-leading sites. Although the surrogate omits the constant terms of these approximations, these constants cancel when comparing objective values at two points. Consequently, applying the fourth-order Taylor remainder bound and defining
\[
d_N=\sup_{k\ge2}\left\|\widehat\theta_N-\check\theta^{(k)}\right\|,
\]
we obtain, uniformly for $\widetilde\theta\in\partial\mathcal B_N$,
\[
\begin{aligned}
&M_{N,3}^L(\widetilde\theta)-M_{N,3}^L(\widehat\theta_N)
\\
\le{}&
M_N(\widetilde\theta)-M_N(\widehat\theta_N)
+
C_NN\left[(\delta_N+d_N)^4+d_N^4\right]
\\
\le{}&
-\frac{c_0N}{4}\delta_N^2
+
C_NN\left[(\delta_N+d_N)^4+d_N^4\right],
\end{aligned}
\]
where $C_N=O_{P_{\theta,\Gamma}}(1)$ follows from the uniform fourth-derivative bound in Condition C4.

Choosing $\delta_N$ such that
\[
\left(
\delta_N+
\sup_{k\ge2}
\left\|
\widehat\theta_N-\check\theta^{(k)}
\right\|
\right)^4
=
o_{P_{\theta,\Gamma}}(\delta_N^2),
\]
the right-hand side is strictly negative with probability tending to one. Thus, with probability tending to one,
\[
\sup_{\widetilde\theta\in\partial\mathcal B_N}
M_{N,3}^L(\widetilde\theta)
<
M_{N,3}^L(\widehat\theta_N).
\]
Since $M_{N,3}^L$ is continuous, it attains its maximum on the compact set $\mathcal B_N$. The preceding strict inequality implies that this maximum is attained in the interior of $\mathcal B_N$, establishing the existence of a statistically relevant local maximizer $\check\theta_{N,3}^L$.

We now consider the asymptotic properties of $\check\theta_{N,3}^{L}$. For notational convenience, we suppress the subscript $3$ on the proposed cubic-surrogate estimator throughout the remainder of this proof, so $\check\theta_{N}^{L}$ is used instead. The analysis proceeds similar to that in~\cite{chendistributed}, but adapted to handle the third-order terms. By the definition of the global surrogate and the first-order condition,
\[
\dot M_{n_1}^{(1)}(\check\theta_{N}^{L})
+
\sum_{k=2}^K
\left[
\dot M_{n_k}^{(k)}(\check\theta^{(k)})
+
\ddot M_{n_k}^{(k)}(\check\theta^{(k)})
\left(
\check\theta_{N}^{L}-\check\theta^{(k)}
\right)
+
\frac{1}{2}
\dddot M_{n_k}^{(k)}(\check\theta^{(k)})
\left[
\left(
\check\theta_{N}^{L}-\check\theta^{(k)}
\right)^{\otimes 2}
\right]
\right]
=0.
\]
Since the pooled-data estimator satisfies
\[
\sum_{k=1}^K
\dot M_{n_k}^{(k)}(\widehat\theta_N)=0,
\]
combining the two first-order conditions gives
\[
\begin{aligned}
0={}&
\dot M_{n_1}^{(1)}(\check\theta_{N}^{L})
-
\dot M_{n_1}^{(1)}(\widehat\theta_N)
\\
&+
\sum_{k=2}^K
\Bigg[
\dot M_{n_k}^{(k)}(\check\theta^{(k)})
-
\dot M_{n_k}^{(k)}(\widehat\theta_N)
+
\ddot M_{n_k}^{(k)}(\check\theta^{(k)})
\left(
\check\theta_{N}^{L}-\check\theta^{(k)}
\right)
\\
&\hspace{4cm}
+
\frac{1}{2}
\dddot M_{n_k}^{(k)}(\check\theta^{(k)})
\left[
\left(
\check\theta_{N}^{L}-\check\theta^{(k)}
\right)^{\otimes 2}
\right]
\Bigg].
\end{aligned}
\]
For each $k\geq 2$, by decomposing
\[
\check\theta_{N}^{L} - \check\theta^{(k)}
=
\left(
\check\theta_{N}^{L} - \widehat\theta_N
\right)
+
\left(
\widehat\theta_N - \check\theta^{(k)}
\right),
\]
we find that this expression is equivalent to

\[
\begin{aligned}
0={}&
\dot M_{n_1}^{(1)}(\check\theta_N^L)
-
\dot M_{n_1}^{(1)}(\widehat\theta_N)
\\
&+
\sum_{k=2}^K
\Bigg[
\ddot M_{n_k}^{(k)}(\check\theta^{(k)})
\left(
\check\theta_N^L-\widehat\theta_N
\right)
+
\dddot M_{n_k}^{(k)}(\check\theta^{(k)})
\left[
\frac{1}{2}\check\theta_N^L
+
\frac{1}{2}\widehat\theta_N
-
\check\theta^{(k)},
\check\theta_N^L-\widehat\theta_N
\right]
\Bigg]
\\
&+
\sum_{k=2}^K
\Bigg[
\dot M_{n_k}^{(k)}(\check\theta^{(k)})
-
\dot M_{n_k}^{(k)}(\widehat\theta_N)
-
\ddot M_{n_k}^{(k)}(\check\theta^{(k)})
\left(
\check\theta^{(k)}-\widehat\theta_N
\right)
+
\frac{1}{2}
\dddot M_{n_k}^{(k)}(\check\theta^{(k)})
\left[
\left(
\check\theta^{(k)}-\widehat\theta_N
\right)^{\otimes2}
\right]
\Bigg].
\end{aligned}
\]

Using the integral form of Taylor's theorem, the previous expression is equivalent to
\begin{equation}
\label{eq:cubic-leading-rate}
\begin{aligned}
&
\frac{1}{N}
\Bigg[
\int_0^1
\ddot M_{n_1}^{(1)}
\left(
\widehat\theta_N
+
u(\check\theta_N^L-\widehat\theta_N)
\right)
\,du
\\
&\qquad
+
\sum_{k=2}^K
\Bigg\{
\ddot M_{n_k}^{(k)}(\check\theta^{(k)})
+
\dddot M_{n_k}^{(k)}(\check\theta^{(k)})
\left[
\frac{1}{2}\check\theta_N^L
+
\frac{1}{2}\widehat\theta_N
-
\check\theta^{(k)}
\right]
\Bigg\}
\Bigg]
\left(
\check\theta_N^L-\widehat\theta_N
\right)
\\
&=
-\frac{1}{N}
\sum_{k=2}^K
\Bigg[
\dot M_{n_k}^{(k)}(\check\theta^{(k)})
-
\dot M_{n_k}^{(k)}(\widehat\theta_N)
-
\ddot M_{n_k}^{(k)}(\check\theta^{(k)})
\left(
\check\theta^{(k)}-\widehat\theta_N
\right)
\\
&\qquad\qquad
+
\frac{1}{2}
\dddot M_{n_k}^{(k)}(\check\theta^{(k)})
\left[
\left(
\check\theta^{(k)}-\widehat\theta_N
\right)^{\otimes2}
\right]
\Bigg].
\end{aligned}
\end{equation}

The coefficient matrix on the left-hand side combines the integrated Hessian of the leading-site criterion with the midpoint Hessians of the non-leading cubic approximations, while the right-hand side consists of the third-order Taylor remainders arising from the approximation of the site-specific scores.

We now consider the expression on the left-hand side of Equation~\eqref{eq:cubic-leading-rate}, and specifically the difference between its coefficient matrix and $-\widetilde I$. We have
\[
\begin{aligned}
&
\frac{1}{N}
\Bigg[
\int_0^1
\ddot M_{n_1}^{(1)}
\left(
\widehat\theta_N
+
u(\check\theta_N^L-\widehat\theta_N)
\right)
\,du
\\
&\qquad
+
\sum_{k=2}^K
\Bigg\{
\ddot M_{n_k}^{(k)}(\check\theta^{(k)})
+
\dddot M_{n_k}^{(k)}(\check\theta^{(k)})
\left[
\frac{1}{2}\check\theta_N^L
+
\frac{1}{2}\widehat\theta_N
-
\check\theta^{(k)}
\right]
\Bigg\}
\Bigg]
+
\widetilde I
\\
={}&
\frac{1}{N}
\left[
\int_0^1
\ddot M_{n_1}^{(1)}
\left(
\widehat\theta_N
+
u(\check\theta_N^L-\widehat\theta_N)
\right)
\,du
-
\ddot M_{n_1}^{(1)}(\theta)
\right]
\\
&+
\frac{1}{N}
\sum_{k=2}^K
\left[
\ddot M_{n_k}^{(k)}(\check\theta^{(k)})
-
\ddot M_{n_k}^{(k)}(\theta)
\right]
\\
&+
\frac{1}{N}
\sum_{k=2}^K
\dddot M_{n_k}^{(k)}(\check\theta^{(k)})
\left[
\frac{1}{2}\check\theta_N^L
+
\frac{1}{2}\widehat\theta_N
-
\check\theta^{(k)}
\right]
\\
&+
\frac{1}{N}
\sum_{k=1}^K
\left[
\ddot M_{n_k}^{(k)}(\theta)
+
n_k\widetilde I_k
\right]
+
\left[
\widetilde I
-
\frac{1}{N}
\sum_{k=1}^K
n_k\widetilde I_k
\right].
\end{aligned}
\]

The first term is controlled by the uniform boundedness of the third
derivatives in Condition C4. Specifically,
\[
\begin{aligned}
&
\left\|
\frac{1}{N}
\left[
\int_0^1
\ddot M_{n_1}^{(1)}
\left(
\widehat\theta_N
+
u(\check\theta_N^L-\widehat\theta_N)
\right)
\,du
-
\ddot M_{n_1}^{(1)}(\theta)
\right]
\right\|
\\
&\qquad\leq
\sup_{\vartheta\in\Theta}
\left\|
\frac{\dddot M_{n_1}^{(1)}(\vartheta)}{n_1}
\right\|
\left\|
\widehat\theta_N-\theta
\right\|
+
\sup_{\vartheta\in\Theta}
\left\|
\frac{\dddot M_{n_1}^{(1)}(\vartheta)}{n_1}
\right\|
\left\|
\check\theta_N^L-\widehat\theta_N
\right\|
=o_{P_{\theta,\Gamma}}(1),
\end{aligned}
\]
where we use the consistency of $\widehat\theta_N$ and $\check\theta_N^L$.

Similarly, for the second term,
\[
\begin{aligned}
&
\left\|
\frac{1}{N}
\sum_{k=2}^K
\left[
\ddot M_{n_k}^{(k)}(\check\theta^{(k)})
-
\ddot M_{n_k}^{(k)}(\theta)
\right]
\right\|
\\
&\qquad\leq
\sup_{k=2,\ldots,K}
\sup_{\vartheta\in\Theta}
\left\|
\frac{\dddot M_{n_k}^{(k)}(\vartheta)}{n_k}
\right\|
\sup_{k=2,\ldots,K}
\left\|
\check\theta^{(k)}-\theta
\right\|
=o_{P_{\theta,\Gamma}}(1).
\end{aligned}
\]

For the third term, which is introduced by the cubic approximation, the
same uniform boundedness condition gives
\[
\begin{aligned}
&
\left\|
\frac{1}{N}
\sum_{k=2}^K
\dddot M_{n_k}^{(k)}(\check\theta^{(k)})
\left[
\frac{1}{2}\check\theta_N^L
+
\frac{1}{2}\widehat\theta_N
-
\check\theta^{(k)}
\right]
\right\|
\\
&\qquad\leq
\sup_{k=2,\ldots,K}
\left\|
\frac{\dddot M_{n_k}^{(k)}(\check\theta^{(k)})}{n_k}
\right\|
\sup_{k=2,\ldots,K}
\left\|
\frac{1}{2}\check\theta_N^L
+
\frac{1}{2}\widehat\theta_N
-
\check\theta^{(k)}
\right\|.
\end{aligned}
\]
The first factor is $O_{P_{\theta,\Gamma}}(1)$ by Condition C4, while
\[
\begin{aligned}
\sup_{k=2,\ldots,K}
\left\|
\frac{1}{2}\check\theta_N^L
+
\frac{1}{2}\widehat\theta_N
-
\check\theta^{(k)}
\right\|
&\leq
\frac{1}{2}\|\check\theta_N^L-\theta\|
+
\frac{1}{2}\|\widehat\theta_N-\theta\|
+
\sup_{k=2,\ldots,K}
\|\check\theta^{(k)}-\theta\|
\\
&=o_{P_{\theta,\Gamma}}(1).
\end{aligned}
\]
Thus, the third term is also
\[
o_{P_{\theta,\Gamma}}(1).
\]

For the fourth term, Condition C4 implies
\[
\begin{aligned}
\left\|
\frac{1}{N}
\sum_{k=1}^K
\left[
\ddot M_{n_k}^{(k)}(\theta)
+
n_k\widetilde I_k
\right]
\right\|
&\leq
\sup_{k=1,\ldots,K}
\left\|
\frac{\ddot M_{n_k}^{(k)}(\theta)}{n_k}
+
\widetilde I_k
\right\|
=o_{P_{\theta,\Gamma}}(1).
\end{aligned}
\]

Finally, by Condition C3,
\[
\widetilde I
-
\frac{1}{N}
\sum_{k=1}^K n_k\widetilde I_k
\longrightarrow 0.
\]

Therefore, combining the five terms,
\[
\begin{aligned}
&
\frac{1}{N}
\Bigg[
\int_0^1
\ddot M_{n_1}^{(1)}
\left(
\widehat\theta_N
+
u(\check\theta_N^L-\widehat\theta_N)
\right)\,du
\\
&\qquad
+
\sum_{k=2}^K
\Bigg\{
\ddot M_{n_k}^{(k)}(\check\theta^{(k)})
+
\dddot M_{n_k}^{(k)}(\check\theta^{(k)})
\left[
\frac{1}{2}\check\theta_N^L
+
\frac{1}{2}\widehat\theta_N
-
\check\theta^{(k)}
\right]
\Bigg\}
\Bigg]
+
\widetilde I
=o_{P_{\theta,\Gamma}}(1),
\end{aligned}
\]
and hence
\[
\frac{1}{N}
\Bigg[
\int_0^1
\ddot M_{n_1}^{(1)}
\left(
\widehat\theta_N
+
u(\check\theta_N^L-\widehat\theta_N)
\right)\,du
+
\sum_{k=2}^K
\Bigg\{
\ddot M_{n_k}^{(k)}(\check\theta^{(k)})
+
\dddot M_{n_k}^{(k)}(\check\theta^{(k)})
\left[
\frac{1}{2}\check\theta_N^L
+
\frac{1}{2}\widehat\theta_N
-
\check\theta^{(k)}
\right]
\Bigg\}
\Bigg]
\overset{P_{\theta,\Gamma}}{\longrightarrow}
-\widetilde I.
\]

We next consider the right-hand side of Equation~\eqref{eq:cubic-leading-rate}, denoted by $R_N^L$. This quantity is the aggregate third-order Taylor remainder obtained by expanding the local gradients around $\check\theta^{(k)}$. By the fourth-order smoothness condition,
\[
\begin{aligned}
\|R_N^L\|
\le{}&
\frac{1}{6N}
\sum_{k=2}^K
\sup_{\vartheta\in\Theta}
\left\|
\ddddot M_{n_k}^{(k)}(\vartheta)
\right\|
\left\|
\check\theta^{(k)}-\widehat\theta_N
\right\|^3
\\
\le{}&
\frac16
\sup_{k=2,\ldots,K}
\sup_{\vartheta\in\Theta}
\left\|
\frac{\ddddot M_{n_k}^{(k)}(\vartheta)}{n_k}
\right\|
\left(
\sup_{k=2,\ldots,K}
\left\|
\check\theta^{(k)}-\widehat\theta_N
\right\|
\right)^3
\\
={}&
o_{P_{\theta,\Gamma}}(N^{-3/4}).
\end{aligned}
\]

Since
\[
\frac{1}{N}
\Bigg[
\int_0^1
\ddot M_{n_1}^{(1)}
\left(
\widehat\theta_N
+
u(\check\theta_N^L-\widehat\theta_N)
\right)\,du
+
\sum_{k=2}^K
\Bigg\{
\ddot M_{n_k}^{(k)}(\check\theta^{(k)})
+
\dddot M_{n_k}^{(k)}(\check\theta^{(k)})
\left[
\frac{1}{2}\check\theta_N^L
+
\frac{1}{2}\widehat\theta_N
-
\check\theta^{(k)}
\right]
\Bigg\}
\Bigg]
\overset{P_{\theta,\Gamma}}{\longrightarrow}
-\widetilde I,
\]
and $\widetilde I$ is positive definite by Condition C3, the coefficient
matrix on the left-hand side of Equation~\eqref{eq:cubic-leading-rate} is
invertible with probability tending to one, with its inverse bounded in
probability. Therefore,
\[
\begin{aligned}
\left\|
\check\theta_N^L-\widehat\theta_N
\right\|
&=
O_{P_{\theta,\Gamma}}
\left(
\sup_{k=2,\ldots,K}
\left\|
\check\theta^{(k)}-\widehat\theta_N
\right\|^3
\right).
\end{aligned}
\]
From the results established above,
\[
\sup_{k=2,\ldots,K}
\left\|
\check\theta^{(k)}-\widehat\theta_N
\right\|
=
o_{P_{\theta,\Gamma}}(N^{-1/4}),
\]
and hence
\[
\boxed{
\check\theta_N^L-\widehat\theta_N
=
o_{P_{\theta,\Gamma}}(N^{-3/4}).
}
\]

\subsubsection{Convergence Rate of the Leading-Site Calibrated Estimator}
\label{subsubsec:proof-calibrated-rate-leading}

We now demonstrate the convergence rate for $\widehat\theta_N^L$. As
above, to simplify notation, we suppress the subscript $3$ throughout the
remainder of the proof. Thus, $\check\theta_N^L$ denotes the cubic
surrogate estimator and $\widehat\theta_N^L$ denotes the corresponding
cubic-calibrated estimator. From the preceding argument,
\[
\check\theta_N^L-\widehat\theta_N
=
o_{P_{\theta,\Gamma}}(N^{-3/4}).
\]
For the calibrated estimator, we additionally invoke the fifth-order smoothness condition stated in C4, which is the direct higher-order analogue of the additional smoothness condition imposed for calibration in~\cite{chendistributed}.

We first establish the existence of a statistically relevant local maximizer of the cubic calibration criterion. As in the surrogate existence argument, we show that the criterion is locally strongly concave and takes strictly smaller values on the boundary of a shrinking neighborhood of the pooled-data estimator.

Let $\Psi_N^L(\vartheta)$ denote the cubic calibration criterion expressed in terms of the parameter $\vartheta$, rather than the displacement from $\check\theta_N^L$. By the fourth- and fifth-order Taylor remainder bounds, together with the previously established surrogate convergence rate, we have
\[
\begin{aligned}
\frac1N
\left\|
\nabla\Psi_N^L(\widehat\theta_N)
\right\|
\le{}&
O_{P_{\theta,\Gamma}}(1)
\left\|
\check\theta_N^L-\widehat\theta_N
\right\|^3
\\
&+
O_{P_{\theta,\Gamma}}(1)
\sup_{k\ge2}
\left\|
\widehat\theta_N-\check\theta^{(k)}
\right\|^4
=
o_{P_{\theta,\Gamma}}(N^{-1}).
\end{aligned}
\]
Here, the first term arises from the Taylor approximation of the exact leading-site score around $\check\theta_N^L$, while the second arises from transporting the non-leading-site scores from their local expansion points.

Moreover, the uniform derivative bounds and localization results imply
\[
-\frac1N
\nabla^2\Psi_N^L(\widehat\theta_N)
=
\widetilde I+o_{P_{\theta,\Gamma}}(1).
\]
Since the Hessian of the cubic calibration criterion is affine and its normalized third derivative is bounded in probability, the criterion is uniformly strongly concave in a sufficiently small neighborhood of $\widehat\theta_N$.

In particular, let $\mathcal B_N=B_{r_N}(\widehat\theta_N)$ with $r_N=N^{-1/2}$, and let $c_0=\lambda_{\min}(\widetilde I)>0$. With probability tending to one, the normalized negative Hessian is bounded below by $c_0I/2$ throughout $\mathcal B_N$. Consequently, for $\vartheta\in\partial\mathcal B_N$,
\[
\begin{aligned}
\Psi_N^L(\vartheta)
-
\Psi_N^L(\widehat\theta_N)
\le{}&
\left\|
\nabla\Psi_N^L(\widehat\theta_N)
\right\|r_N
-
\frac{c_0N}{4}r_N^2
\\
={}&
o_{P_{\theta,\Gamma}}(N^{-1/2})
-
\frac{c_0}{4}
<0
\end{aligned}
\]
with probability tending to one. By continuity and compactness, the calibration criterion therefore attains its maximum in the interior of $\mathcal B_N$, establishing the existence of a statistically relevant local maximizer $\widehat\theta_N^L$.

To study the difference between the calibrated estimator and the pooled-data
estimator, we decompose the transported score at the surrogate estimator as
\[
\widehat G_N^L(\check\theta_N^L)
-
\dot M_N(\widehat\theta_N)
=
\underbrace{
\left[
\widehat G_N^L(\check\theta_N^L)
-
\widehat G_N^L(\widehat\theta_N)
\right]
}_{\text{change in the transported score}}
+
\underbrace{
\left[
\widehat G_N^L(\widehat\theta_N)
-
\dot M_N(\widehat\theta_N)
\right]
}_{\text{transport error}}
.
\]

We first compare the transported score with the true pooled score at
$\widehat\theta_N$. By the definition of the cubic calibration surrogate,
the aggregated score evaluated at $\widehat\theta_N$ is
\[
\begin{aligned}
\widehat G_N^L(\widehat\theta_N)
={}&
\dot M_{n_1}^{(1)}(\widehat\theta_N)
\\
&+
\sum_{k=2}^K
\Bigg[
\dot M_{n_k}^{(k)}(\check\theta^{(k)})
+
\ddot M_{n_k}^{(k)}(\check\theta^{(k)})
\left(
\widehat\theta_N-\check\theta^{(k)}
\right)
\\
&\qquad\qquad
+
\frac{1}{2}
\dddot M_{n_k}^{(k)}(\check\theta^{(k)})
\left[
\left(
\widehat\theta_N-\check\theta^{(k)}
\right)^{\otimes2}
\right]
\\
&\qquad\qquad
+
\frac{1}{6}
\ddddot M_{n_k}^{(k)}(\check\theta^{(k)})
\left[
\left(
\widehat\theta_N-\check\theta^{(k)}
\right)^{\otimes3}
\right]
\Bigg].
\end{aligned}
\]
On the other hand, the pooled-data estimator satisfies
\[
\dot M_N(\widehat\theta_N)
=
\dot M_{n_1}^{(1)}(\widehat\theta_N)
+
\sum_{k=2}^K
\dot M_{n_k}^{(k)}(\widehat\theta_N)
=
0.
\]
Therefore,
\[
\begin{aligned}
\widehat G_N^L(\widehat\theta_N)
&=
\widehat G_N^L(\widehat\theta_N)
-
\dot M_N(\widehat\theta_N)
\\
&=
\sum_{k=2}^K
\Bigg[
\dot M_{n_k}^{(k)}(\check\theta^{(k)})
+
\ddot M_{n_k}^{(k)}(\check\theta^{(k)})
\left(
\widehat\theta_N-\check\theta^{(k)}
\right)
\\
&\qquad\qquad
+
\frac{1}{2}
\dddot M_{n_k}^{(k)}(\check\theta^{(k)})
\left[
\left(
\widehat\theta_N-\check\theta^{(k)}
\right)^{\otimes2}
\right]
\\
&\qquad\qquad
+
\frac{1}{6}
\ddddot M_{n_k}^{(k)}(\check\theta^{(k)})
\left[
\left(
\widehat\theta_N-\check\theta^{(k)}
\right)^{\otimes3}
\right]
-
\dot M_{n_k}^{(k)}(\widehat\theta_N)
\Bigg].
\end{aligned}
\]
Observe that the terms inside the sum are precisely the derivatives of the fourth-order Taylor approximations to the local criteria $M_{n_k}^{(k)}$ around $\check\theta^{(k)}$. Since $\dot M_N(\widehat\theta_N)=0$, the leading-site terms cancel, leaving only the higher-order Taylor remainder terms. Thus,
\begin{equation}
\label{eq:calibration-transport-error}
\begin{aligned}
&
\widehat G_N^L(\widehat\theta_N)
-
\dot M_N(\widehat\theta_N)
\\
&=
-\frac{1}{6}
\sum_{k=2}^K
\int_0^1
(1-u)^3
\overset{5}{M}_{n_k}^{(k)}
\left(
\check\theta^{(k)}
+
u\left(
\widehat\theta_N-\check\theta^{(k)}
\right)
\right)
\,du
\left[
\left(
\widehat\theta_N-\check\theta^{(k)}
\right)^{\otimes4}
\right].
\end{aligned}
\end{equation}

We next compare the transported score evaluated at the surrogate estimator
with the same quantity evaluated at the pooled-data estimator. Recall that
\[
\widehat G_N^L(\vartheta)
=
\dot M_{n_1}^{(1)}(\vartheta)
+
\sum_{k=2}^K
\widehat G_{k}^L(\vartheta).
\]
For the leading site, the score is evaluated exactly. By the fundamental
theorem of calculus,
\[
\begin{aligned}
\dot M_{n_1}^{(1)}(\check\theta_N^L)
-
\dot M_{n_1}^{(1)}(\widehat\theta_N)
={}&
\ddot M_{n_1}^{(1)}(\widehat\theta_N)
\left(
\check\theta_N^L-\widehat\theta_N
\right)
\\
&+
\frac{1}{2}
\dddot M_{n_1}^{(1)}(\widehat\theta_N)
\left[
\left(
\check\theta_N^L-\widehat\theta_N
\right)^{\otimes2}
\right]
\\
&+
\Bigg[
\int_0^1\int_0^1\int_0^1
u^2v\,
\ddddot M_{n_1}^{(1)}
\left(
\widehat\theta_N
+
uvw
\left(
\check\theta_N^L-\widehat\theta_N
\right)
\right)
\,dw\,dv\,du
\Bigg]
\left[
\left(
\check\theta_N^L-\widehat\theta_N
\right)^{\otimes3}
\right].
\end{aligned}
\]
For each non-leading site $k\geq2$, the transported score is
\[
\begin{aligned}
\widehat G_{k}^L(\vartheta)
={}&
\dot M_{n_k}^{(k)}(\check\theta^{(k)})
+
\ddot M_{n_k}^{(k)}(\check\theta^{(k)})
\left(
\vartheta-\check\theta^{(k)}
\right)
\\
&+
\frac{1}{2}
\dddot M_{n_k}^{(k)}(\check\theta^{(k)})
\left[
\left(
\vartheta-\check\theta^{(k)}
\right)^{\otimes2}
\right]
\\
&+
\frac{1}{6}
\ddddot M_{n_k}^{(k)}(\check\theta^{(k)})
\left[
\left(
\vartheta-\check\theta^{(k)}
\right)^{\otimes3}
\right].
\end{aligned}
\]
We are interested in
\[
\begin{aligned}
\widehat G^L_k(\check\theta_N^L)
-
\widehat G^L_k(\widehat\theta_N)
={}&
\dot M_{n_k}^{(k)}(\check\theta^{(k)})
-
\dot M_{n_k}^{(k)}(\check\theta^{(k)})
\\
&+
\ddot M_{n_k}^{(k)}(\check\theta^{(k)})
\left(
\check\theta_N^L-\check\theta^{(k)}
\right)
-
\ddot M_{n_k}^{(k)}(\check\theta^{(k)})
\left(
\widehat\theta_N-\check\theta^{(k)}
\right)
\\
&+
\frac12
\dddot M_{n_k}^{(k)}(\check\theta^{(k)})
\left[
\left(
\check\theta_N^L-\check\theta^{(k)}
\right)^{\otimes2}
\right]
-
\frac12
\dddot M_{n_k}^{(k)}(\check\theta^{(k)})
\left[
\left(
\widehat\theta_N-\check\theta^{(k)}
\right)^{\otimes2}
\right]
\\
&+
\frac16
\ddddot M_{n_k}^{(k)}(\check\theta^{(k)})
\left[
\left(
\check\theta_N^L-\check\theta^{(k)}
\right)^{\otimes3}
\right]
-
\frac16
\ddddot M_{n_k}^{(k)}(\check\theta^{(k)})
\left[
\left(
\widehat\theta_N-\check\theta^{(k)}
\right)^{\otimes3}
\right]
\\[4pt]
={}&
0
\\
&+
\ddot M_{n_k}^{(k)}(\check\theta^{(k)})
\left(
\check\theta_N^L-\widehat\theta_N
\right)
\\
&+
\dddot M_{n_k}^{(k)}(\check\theta^{(k)})
\left[
\widehat\theta_N-\check\theta^{(k)},
\check\theta_N^L-\widehat\theta_N
\right]
+
\frac12
\dddot M_{n_k}^{(k)}(\check\theta^{(k)})
\left[
\left(
\check\theta_N^L-\widehat\theta_N
\right)^{\otimes2}
\right]
\\
&+
\frac12
\ddddot M_{n_k}^{(k)}(\check\theta^{(k)})
\left[
\left(
\widehat\theta_N-\check\theta^{(k)}
\right)^{\otimes2},
\check\theta_N^L-\widehat\theta_N
\right]
\\
&\qquad+
\frac12
\ddddot M_{n_k}^{(k)}(\check\theta^{(k)})
\left[
\widehat\theta_N-\check\theta^{(k)},
\left(
\check\theta_N^L-\widehat\theta_N
\right)^{\otimes2}
\right]
+
\frac16
\ddddot M_{n_k}^{(k)}(\check\theta^{(k)})
\left[
\left(
\check\theta_N^L-\widehat\theta_N
\right)^{\otimes3}
\right].
\end{aligned}
\]

Finally, we can consider

\[
\begin{aligned}
\widehat G_N^L(\check\theta_N^L)
-
\widehat G_N^L(\widehat\theta_N)
={}&
\Bigg[
\ddot M_{n_1}^{(1)}(\widehat\theta_N)
+
\sum_{k=2}^K
\Bigg\{
\ddot M_{n_k}^{(k)}(\check\theta^{(k)})
+
\dddot M_{n_k}^{(k)}(\check\theta^{(k)})
\left[
\widehat\theta_N-\check\theta^{(k)}
\right]
\\
&\qquad\qquad\qquad
+
\frac12
\ddddot M_{n_k}^{(k)}(\check\theta^{(k)})
\left[
\left(
\widehat\theta_N-\check\theta^{(k)}
\right)^{\otimes2}
\right]
\Bigg\}
\Bigg]
\left(
\check\theta_N^L-\widehat\theta_N
\right)
\\
&+
\frac12
\Bigg[
\dddot M_{n_1}^{(1)}(\widehat\theta_N)
+
\sum_{k=2}^K
\Bigg\{
\dddot M_{n_k}^{(k)}(\check\theta^{(k)})
\\
&\qquad\qquad\qquad
+
\ddddot M_{n_k}^{(k)}(\check\theta^{(k)})
\left[
\widehat\theta_N-\check\theta^{(k)}
\right]
\Bigg\}
\Bigg]
\left[
\left(
\check\theta_N^L-\widehat\theta_N
\right)^{\otimes2}
\right]
\\
&+
\frac16
\sum_{k=2}^K
\ddddot M_{n_k}^{(k)}(\check\theta^{(k)})
\left[
\left(
\check\theta_N^L-\widehat\theta_N
\right)^{\otimes3}
\right]
\\
&+
\Bigg[
\int_0^1
\int_0^1
\int_0^1
u^2v\,
\ddddot M_{n_1}^{(1)}
\left(
\widehat\theta_N
+
uvw
\left(
\check\theta_N^L-\widehat\theta_N
\right)
\right)
\,dw\,dv\,du
\Bigg]
\left[
\left(
\check\theta_N^L-\widehat\theta_N
\right)^{\otimes3}
\right].
\end{aligned}
\]
Thus, the difference in transported scores can be organized as
\begin{equation}
\label{eq:calibration-transported-score-increment}
\begin{aligned}
\widehat G_N^L(\check\theta_N^L)
-
\widehat G_N^L(\widehat\theta_N)
={}&
\widehat H_N^L(\widehat\theta_N)
\left(
\check\theta_N^L-\widehat\theta_N
\right)
\\
&+
\frac{1}{2}
\widehat T_N^L(\widehat\theta_N)
\left[
\left(
\check\theta_N^L-\widehat\theta_N
\right)^{\otimes2}
\right]
\\
&+
\frac{1}{6}
\sum_{k=2}^K
\ddddot M_{n_k}^{(k)}(\check\theta^{(k)})
\left[
\left(
\check\theta_N^L-\widehat\theta_N
\right)^{\otimes3}
\right]
\\
&+
\Bigg[
\int_0^1
\int_0^1
\int_0^1
u^2v\,
\ddddot M_{n_1}^{(1)}
\left(
\widehat\theta_N
+
uvw
\left(
\check\theta_N^L-\widehat\theta_N
\right)
\right)
\,dw\,dv\,du
\Bigg]
\left[
\left(
\check\theta_N^L-\widehat\theta_N
\right)^{\otimes3}
\right].
\end{aligned}
\end{equation}

Having obtained \eqref{eq:calibration-transport-error} and
\eqref{eq:calibration-transported-score-increment}, we return to
$\widehat G_N^L(\check\theta_N^L)$, which is the score term appearing in the
first-order condition for the cubic calibration criterion. Since
$\dot M_N(\widehat\theta_N)=0$, we have
\[
\widehat G_N^L(\check\theta_N^L)
=
\left[
\widehat G_N^L(\check\theta_N^L)
-
\widehat G_N^L(\widehat\theta_N)
\right]
+
\left[
\widehat G_N^L(\widehat\theta_N)
-
\dot M_N(\widehat\theta_N)
\right].
\]
Substituting \eqref{eq:calibration-transported-score-increment} and
\eqref{eq:calibration-transport-error} gives
\begin{equation}
\label{eq:combined-gradient-representation}
\begin{aligned}
\widehat G_N^L(\check\theta_N^L)
={}&
\widehat H_N^L(\widehat\theta_N)
\left(
\check\theta_N^L-\widehat\theta_N
\right)
\\
&+
\frac12
\widehat T_N^L(\widehat\theta_N)
\left[
\left(
\check\theta_N^L-\widehat\theta_N
\right)^{\otimes2}
\right]
\\
&+
\frac16
\sum_{k=2}^K
\ddddot M_{n_k}^{(k)}(\check\theta^{(k)})
\left[
\left(
\check\theta_N^L-\widehat\theta_N
\right)^{\otimes3}
\right]
\\
&+
\Bigg[
\int_0^1
\int_0^1
\int_0^1
u^2v\,
\ddddot M_{n_1}^{(1)}
\left(
\widehat\theta_N
+
uvw
\left(
\check\theta_N^L-\widehat\theta_N
\right)
\right)
\,dw\,dv\,du
\Bigg]\times
\left[
\left(
\check\theta_N^L-\widehat\theta_N
\right)^{\otimes3}
\right]
\\
&-
\frac{1}{6}
\sum_{k=2}^K
\int_0^1
(1-u)^3
\overset{5}{M}_{n_k}^{(k)}
\left(
\check\theta^{(k)}
+
u\left(
\widehat\theta_N-\check\theta^{(k)}
\right)
\right)
\,du
\left[
\left(
\widehat\theta_N-\check\theta^{(k)}
\right)^{\otimes4}
\right].
\end{aligned}
\end{equation}

Having established the existence of a statistically relevant local maximizer of the cubic calibration criterion, we now invoke its first-order condition,

\[
\widehat G_N^L(\check\theta_N^L)
+
\widehat H_N^L(\check\theta_N^L)
\left(
\widehat\theta_N^L-\check\theta_N^L
\right)
+
\frac{1}{2}
\widehat T_N^L(\check\theta_N^L)
\left[
\left(
\widehat\theta_N^L-\check\theta_N^L
\right)^{\otimes2}
\right]
=0.
\]
Add and subtract the cubic calibration gradient evaluated at
$\widehat\theta_N$, namely
\[
\widehat G_N^L(\check\theta_N^L)
+
\widehat H_N^L(\check\theta_N^L)
\left(
\widehat\theta_N-\check\theta_N^L
\right)
+
\frac{1}{2}
\widehat T_N^L(\check\theta_N^L)
\left[
\left(
\widehat\theta_N-\check\theta_N^L
\right)^{\otimes2}
\right].
\]
Using
\[
\begin{aligned}
&
\frac{1}{2}
\Bigg[
\left(
\widehat\theta_N^L-\check\theta_N^L
\right)^{\otimes2}
-
\left(
\widehat\theta_N-\check\theta_N^L
\right)^{\otimes2}
\Bigg]=
\left[
\frac{
\widehat\theta_N^L+\widehat\theta_N
}{2}
-\check\theta_N^L,\,
\widehat\theta_N^L-\widehat\theta_N
\right],
\end{aligned}
\]
we obtain
\[
\begin{aligned}
&
\Bigg[
\widehat H_N^L(\check\theta_N^L)
+
\widehat T_N^L(\check\theta_N^L)
\left[
\frac{
\widehat\theta_N^L+\widehat\theta_N
}{2}
-\check\theta_N^L,\cdot
\right]
\Bigg]
\left(
\widehat\theta_N^L-\widehat\theta_N
\right)
\\
={}&
-\Bigg[
\widehat G_N^L(\check\theta_N^L)
+
\widehat H_N^L(\check\theta_N^L)
\left(
\widehat\theta_N-\check\theta_N^L
\right)+
\frac{1}{2}
\widehat T_N^L(\check\theta_N^L)
\left[
\left(
\widehat\theta_N-\check\theta_N^L
\right)^{\otimes2}
\right]
\Bigg].
\end{aligned}
\]
We now substitute the expansion for
$\widehat G_N^L(\check\theta_N^L)$ from
\eqref{eq:combined-gradient-representation}.

\[
\begin{aligned}
&
\Bigg[
\widehat H_N^L(\check\theta_N^L)
+
\widehat T_N^L(\check\theta_N^L)
\left[
\frac{\widehat\theta_N^L+\widehat\theta_N}{2}
-\check\theta_N^L,\cdot
\right]
\Bigg]
\left(
\widehat\theta_N^L-\widehat\theta_N
\right)
\\
={}&
-\Bigg\{
\left[
\widehat H_N^L(\widehat\theta_N)
-
\widehat H_N^L(\check\theta_N^L)
\right]
\left(
\check\theta_N^L-\widehat\theta_N
\right)
\\
&\qquad+
\frac12
\left[
\widehat T_N^L(\widehat\theta_N)
+
\widehat T_N^L(\check\theta_N^L)
\right]
\left[
\left(
\check\theta_N^L-\widehat\theta_N
\right)^{\otimes2}
\right]
\\
&\qquad+
\frac16
\sum_{k=2}^K
\ddddot M_{n_k}^{(k)}(\check\theta^{(k)})
\left[
\left(
\check\theta_N^L-\widehat\theta_N
\right)^{\otimes3}
\right]
\\
&\qquad+
\Bigg[
\int_0^1
\int_0^1
\int_0^1
u^2v\,
\ddddot M_{n_1}^{(1)}
\left(
\widehat\theta_N
+
uvw
\left(
\check\theta_N^L-\widehat\theta_N
\right)
\right)
\,dw\,dv\,du
\Bigg]
\left[
\left(
\check\theta_N^L-\widehat\theta_N
\right)^{\otimes3}
\right]
\\
&\qquad-
\frac16
\sum_{k=2}^K
\int_0^1
(1-u)^3
\overset{5}{M}_{n_k}^{(k)}
\left(
\check\theta^{(k)}
+
u\left(
\widehat\theta_N-\check\theta^{(k)}
\right)
\right)
\,du
\left[
\left(
\widehat\theta_N-\check\theta^{(k)}
\right)^{\otimes4}
\right]
\Bigg\}.
\end{aligned}
\]

After dividing by $N$, each term on the right-hand side can be bounded using the corresponding derivative condition. The first term is controlled by the uniform bound on the third derivative, the second by the uniform bound on the third derivative, the cubic terms by the fourth derivative, and the final transport remainder by the fifth derivative condition.
Using the established rates for $\check\theta_N^L-\widehat\theta_N$ and $\widehat\theta_N-\check\theta^{(k)}$, all terms on the right-hand side are $o_{P_{\theta,\Gamma}}(N^{-1})$.

\[
\begin{aligned}
&
\frac{1}{N}
\Bigg\|
\Bigg[
\widehat H_N^L(\check\theta_N^L)
+
\widehat T_N^L(\check\theta_N^L)
\left[
\frac{
\widehat\theta_N^L+\widehat\theta_N
}{2}
-\check\theta_N^L,\cdot
\right]
\Bigg]
\left(
\widehat\theta_N^L-\widehat\theta_N
\right)
\Bigg\|
\\
&\le{}
\sup_{\widetilde\theta}
\left\|
\frac{\widehat T_N^L(\widetilde\theta)}{N}
\right\|
\left\|
\check\theta_N^L-\widehat\theta_N
\right\|^2
\\
&\qquad+
\frac12
\sup_{\widetilde\theta}
\left\|
\frac{\widehat T_N^L(\widetilde\theta)}{N}
\right\|
\left\|
\check\theta_N^L-\widehat\theta_N
\right\|^2
\\
&\qquad+
\frac16
\sup_{\widetilde\theta}
\left\|
\frac{\ddddot M_N^L(\widetilde\theta)}{N}
\right\|
\left\|
\check\theta_N^L-\widehat\theta_N
\right\|^3
\\
&\qquad+
\sup_{\widetilde\theta}
\left\|
\frac{\ddddot M_N^L(\widetilde\theta)}{N}
\right\|
\left\|
\check\theta_N^L-\widehat\theta_N
\right\|^3
\\
&\qquad+
\sup_{\widetilde\theta}
\left\|
\frac{\overset{5}{M}_{N}^{(L)}(\widetilde\theta)}{N}
\right\|
\sup_k
\left\|
\widehat\theta_N-\check\theta^{(k)}
\right\|^4
\\[2mm]
&={}
O_{P_{\theta,\Gamma}}(1)
\left\|
\check\theta_N^L-\widehat\theta_N
\right\|^2
+
\frac12
O_{P_{\theta,\Gamma}}(1)
\left\|
\check\theta_N^L-\widehat\theta_N
\right\|^2
+
\frac16
O_{P_{\theta,\Gamma}}(1)
\left\|
\check\theta_N^L-\widehat\theta_N
\right\|^3
\\
&\qquad+
O_{P_{\theta,\Gamma}}(1)
\left\|
\check\theta_N^L-\widehat\theta_N
\right\|^3+
O_{P_{\theta,\Gamma}}(1)
\sup_k
\left\|
\widehat\theta_N-\check\theta^{(k)}
\right\|^4
\\[2mm]
&={}
o_{P_{\theta,\Gamma}}(N^{-3/2})+
\frac12
o_{P_{\theta,\Gamma}}(N^{-3/2})+
\frac16
o_{P_{\theta,\Gamma}}(N^{-9/4})+
o_{P_{\theta,\Gamma}}(N^{-9/4})+
o_{P_{\theta,\Gamma}}(N^{-1})
\\[2mm]
&={}
o_{P_{\theta,\Gamma}}(N^{-1}).
\end{aligned}
\]
It remains to establish that the matrix multiplying $\widehat\theta_N^L-\widehat\theta_N$ is uniformly nonsingular. Define
\[
\mathcal H_N^{\mathrm{eff}}
:=
\widehat H_N^L(\check\theta_N^L)
+
\widehat T_N^L(\check\theta_N^L)
\left[
\frac{
\widehat\theta_N^L+\widehat\theta_N
}{2}
-\check\theta_N^L,\cdot
\right].
\]
Since the Hessian of the cubic calibration criterion is affine in its
argument,
\[
\mathcal H_N^{\mathrm{eff}}
=
\frac12
\left[
\mathcal H_N^L(\widehat\theta_N)
+
\mathcal H_N^L(\widehat\theta_N^L)
\right],
\]
where
\[
\mathcal H_N^L(\vartheta)
=
\widehat H_N^L(\check\theta_N^L)
+
\widehat T_N^L(\check\theta_N^L)
\left[
\vartheta-\check\theta_N^L,\cdot
\right]
\]
denotes the Hessian of the cubic calibration criterion.

By C3, the limiting information matrix $\widetilde I$ is positive
definite. Moreover, by the preceding uniform convergence and
localization results,
\[
-\frac1N
\mathcal H_N^L(\widehat\theta_N)
=
\widetilde I
+
o_{P_{\theta,\Gamma}}(1).
\]
Thus, if
$c_0=\lambda_{\min}(\widetilde I)>0$, then with probability tending to
one,
\[
-\frac1N
\mathcal H_N^L(\widehat\theta_N)
\succeq
\frac{c_0}{2}I.
\]
On the other hand, since $\widehat\theta_N^L$ is a local maximizer of
the cubic calibration criterion, the second-order necessary condition
gives
\[
-\mathcal H_N^L(\widehat\theta_N^L)\succeq0.
\]
Consequently,
\[
-\frac1N\mathcal H_N^{\mathrm{eff}}
=
-\frac1{2N}
\left[
\mathcal H_N^L(\widehat\theta_N)
+
\mathcal H_N^L(\widehat\theta_N^L)
\right]
\succeq
\frac{c_0}{4}I
\]
with probability tending to one. Hence
\[
\left\|
\left(
\frac1N\mathcal H_N^{\mathrm{eff}}
\right)^{-1}
\right\|
=
O_{P_{\theta,\Gamma}}(1).
\]
Combining this with the preceding bound yields
\[
\left\|
\widehat\theta_N^L-\widehat\theta_N
\right\|
=
o_{P_{\theta,\Gamma}}(N^{-1}).
\]

\subsection{Convergence Rate of the Central-Server Cubic Surrogate Estimator}
\label{subsec:proof-cubic-consistency-central}

Our goal is to establish the convergence rate of the central-server cubic surrogate estimator $\check\theta_{N,3}^{C}$. As in the leading-site proof, we use the pooled-data asymptotic result of Theorem 4.1 and the uniform localization results established in the proof of Theorem 4.2 of~\cite{chendistributed}. In particular, as established in Section~\ref{sec:supp-proof}, the pooled-data estimator satisfies
$$\widehat\theta_N-\theta=O_{P_{\theta,\Gamma}}(N^{-1/2}),$$
while the central-site expansion points satisfy
$$\sup_{k=1,\ldots,K}\|\widetilde\theta_k^C-\theta\|=o_{P_{\theta,\Gamma}}(N^{-1/4}).$$

\subsubsection{Convergence Rate of the Central-Site Surrogate Estimator}
\label{subsubsec:proof-surrogate-rate-central}

We first establish the existence and localization of a statistically relevant local maximizer of the central-server cubic surrogate criterion, denoted by $\check\theta_{N,3}^{C}$. Unlike the leading-site construction, the central-site surrogate does not retain any site-specific criterion exactly, but instead replaces each local criterion with its cubic Taylor approximation around the corresponding local estimator $\widetilde\theta_k^C=\widehat\theta_{n_k}^{(k)}$. Nevertheless, the argument proceeds identically to that of Section~\ref{subsubsec:proof-surrogate-rate-leading}: the separate leading-site contribution is removed, and each summation over the approximated sites is extended from $k=2,\ldots,K$ to $k=1,\ldots,K$.

In particular, the difference between the central-server cubic surrogate Hessian and the pooled Hessian is
\[
\ddot M_{N,3}^{C}(\widetilde\theta)
-
\ddot M_N(\widetilde\theta)
=
-\sum_{k=1}^K
\int_0^1
(1-u)
\ddddot M_{n_k}^{(k)}
\left(
\widetilde\theta_k^C
+
u\left(
\widetilde\theta-\widetilde\theta_k^C
\right)
\right)
\,du
\left[
\left(
\widetilde\theta-\widetilde\theta_k^C
\right)^{\otimes2}
\right].
\]

Similarly, restoring the constant terms omitted from the cubic Taylor approximations, the objective-level approximation error satisfies
\[
\left|
M_{N,3}^{C}(\widetilde\theta)
+
\sum_{k=1}^K
M_{n_k}^{(k)}(\widetilde\theta_k^C)
-
M_N(\widetilde\theta)
\right|
\le
C_N
\sum_{k=1}^K
n_k
\left\|
\widetilde\theta-\widetilde\theta_k^C
\right\|^4,
\]
where $C_N=O_{P_{\theta,\Gamma}}(1)$. 
Since the additional terms are constant with respect to $\widetilde\theta$, they do not affect the maximizer or the comparison of objective values. Let $d_N^C=\sup_{k=1,\ldots,K}\|\widehat\theta_N-\widetilde\theta_k^C\|$. Repeating the boundary comparison from Section~\ref{subsubsec:proof-surrogate-rate-leading}, for $\widetilde\theta\in\partial B_{\delta_N}(\widehat\theta_N)$ we obtain
\[
\begin{aligned}
&M_{N,3}^C(\widetilde\theta)-M_{N,3}^C(\widehat\theta_N)
\\
\le{}&
-\frac{c_0N}{4}\delta_N^2
+
C_NN\left[(\delta_N+d_N^C)^4+(d_N^C)^4\right].
\end{aligned}
\]
Since $d_N^C=o_{P_{\theta,\Gamma}}(N^{-1/4})$, we may choose $\delta_N\to0$ such that $(\delta_N+d_N^C)^4=o_{P_{\theta,\Gamma}}(\delta_N^2)$. The right-hand side is then strictly negative with probability tending to one. Together with the local strong concavity established by the preceding Hessian comparison, this establishes the existence of a statistically relevant local maximizer $\check\theta_{N,3}^{C}$ in a shrinking neighborhood of $\widehat\theta_N$.

We now establish the convergence rate of $\check\theta_{N,3}^{C}$. As in the leading-site case, we compare its first-order condition with the pooled-data first-order condition $\dot M_N(\widehat\theta_N)=0$. The Hessian of the central-server cubic surrogate is
\[
\ddot M_{N,3}^{C}(\vartheta)
=
\sum_{k=1}^K
\left[
\ddot M_{n_k}^{(k)}(\widetilde\theta_k^C)
+
\dddot M_{n_k}^{(k)}(\widetilde\theta_k^C)
\left[
\vartheta-\widetilde\theta_k^C
\right]
\right].
\]
Since this Hessian is affine in $\vartheta$, the coefficient multiplying $\check\theta_{N,3}^{C}-\widehat\theta_N$ in the difference of the two estimating equations is the surrogate Hessian evaluated at their midpoint, namely
\[
\ddot M_{N,3}^{C}
\left(
\frac{\check\theta_{N,3}^{C}+\widehat\theta_N}{2}
\right).
\]

The first-order condition for $\check\theta_{N,3}^{C}$ is obtained by summing the cubic Taylor approximations to the site-specific scores, while the pooled-data estimator satisfies $\dot M_N(\widehat\theta_N)=0$. Subtracting these two estimating equations and collecting the terms multiplying $\check\theta_{N,3}^{C}-\widehat\theta_N$ therefore gives
\[
\begin{aligned}
&
\frac{1}{N}
\ddot M_{N,3}^{C}
\left(
\frac{\check\theta_{N,3}^{C}+\widehat\theta_N}{2}
\right)
\left(
\check\theta_{N,3}^{C}-\widehat\theta_N
\right)
\\
={}&
\frac{1}{2N}
\sum_{k=1}^K
\int_0^1
(1-u)^2
\ddddot M_{n_k}^{(k)}
\left(
\widetilde\theta_k^C
+
u\left(
\widehat\theta_N-\widetilde\theta_k^C
\right)
\right)
\,du
\left[
\left(
\widehat\theta_N-\widetilde\theta_k^C
\right)^{\otimes3}
\right].
\end{aligned}
\]
The right-hand side is precisely the aggregate third-order remainder arising from replacing each site-specific score by its quadratic Taylor approximation around $\widetilde\theta_k^C$. Applying the uniform fourth-derivative bound yields
\[
\begin{aligned}
&
\frac{1}{N}
\left\|
\ddot M_{N,3}^{C}
\left(
\frac{\check\theta_{N,3}^{C}+\widehat\theta_N}{2}
\right)
\left(
\check\theta_{N,3}^{C}-\widehat\theta_N
\right)
\right\|
\\
\le{}&
O_{P_{\theta,\Gamma}}(1)
\sup_{k=1,\ldots,K}
\left\|
\widehat\theta_N-\widetilde\theta_k^C
\right\|^3
=
o_{P_{\theta,\Gamma}}(N^{-3/4}),
\end{aligned}
\]
where the final equality follows from the uniform localization of the site-specific expansion points established above. Moreover, since the midpoint lies within the neighborhood where local strict concavity has already been established,
\[
\left\|
\left[
\frac{1}{N}
\ddot M_{N,3}^{C}
\left(
\frac{\check\theta_{N,3}^{C}+\widehat\theta_N}{2}
\right)
\right]^{-1}
\right\|
=
O_{P_{\theta,\Gamma}}(1).
\]
Multiplying by this uniformly bounded inverse therefore gives
\[
\check\theta_{N,3}^{C}-\widehat\theta_N
=
o_{P_{\theta,\Gamma}}(N^{-3/4}).
\]

\subsubsection{Convergence Rate of the Central-Server Calibrated Estimator}
\label{subsubsec:proof-calibrated-rate-central}

We finally establish the convergence rate of the central-server calibrated estimator $\widehat\theta_{N,3}^{C}$.

We first establish the existence of a statistically relevant local maximizer of the central-server cubic calibration criterion. The argument follows the existence proof in Section~\ref{subsubsec:proof-calibrated-rate-leading}, with the central-server expansion points replacing the leading-site mini-pooled estimators and all site-specific remainder sums extending over $k=1,\ldots,K$. Let $\Psi_N^C(\vartheta)$ denote the central-server cubic calibration criterion expressed in terms of the parameter $\vartheta$, and define $d_N^C=\sup_{k=1,\ldots,K}\|\widehat\theta_N-\widetilde\theta_k^C\|$. By the uniform derivative bounds and the previously established surrogate convergence rate,
\[
\begin{aligned}
\frac1N
\left\|
\nabla\Psi_N^C(\widehat\theta_N)
\right\|
\le{}&
O_{P_{\theta,\Gamma}}(1)
\left(
\left\|
\check\theta_{N,3}^C-\widehat\theta_N
\right\|^3
+
(d_N^C)^4
\right)
\\
={}&
o_{P_{\theta,\Gamma}}(N^{-1}).
\end{aligned}
\]
Moreover, the uniform convergence and localization results imply
\[
-\frac1N
\nabla^2\Psi_N^C(\widehat\theta_N)
=
\widetilde I+o_{P_{\theta,\Gamma}}(1).
\]
Since the Hessian of the cubic calibration criterion is affine and its normalized third derivative is bounded in probability, the criterion remains uniformly strongly concave in a sufficiently small neighborhood of $\widehat\theta_N$. Taking $\mathcal B_N=B_{N^{-1/2}}(\widehat\theta_N)$ and $c_0=\lambda_{\min}(\widetilde I)>0$, we obtain, uniformly for $\vartheta\in\partial\mathcal B_N$,
\[
\begin{aligned}
\Psi_N^C(\vartheta)
-
\Psi_N^C(\widehat\theta_N)
\le{}&
N^{-1/2}
\left\|
\nabla\Psi_N^C(\widehat\theta_N)
\right\|
-
\frac{c_0}{4}
\\
={}&
o_{P_{\theta,\Gamma}}(N^{-1/2})
-
\frac{c_0}{4}
<0
\end{aligned}
\]
with probability tending to one. By continuity and compactness, the calibration criterion therefore attains its maximum in the interior of $\mathcal B_N$, establishing the existence of a statistically relevant local maximizer $\widehat\theta_{N,3}^C$.

We now repeat the argument in Section~\ref{subsubsec:proof-calibrated-rate-leading}. As in the preceding central-server surrogate argument, there is no separately retained leading-site contribution, and all site-specific Taylor and transport remainder sums therefore run from $k=1$ to $K$. The midpoint decomposition of the cubic calibration first-order condition is otherwise unchanged. In particular, the calibration error $\widehat\theta_{N,3}^{C}-\widehat\theta_N$ is multiplied by the average of the Hessians of the central-site cubic calibration criterion at $\widehat\theta_N$ and $\widehat\theta_{N,3}^{C}$, while all remaining terms depend only on the already-controlled surrogate error $\check\theta_{N,3}^{C}-\widehat\theta_N$ and the corresponding higher-order Taylor remainders.

Using
\[
\check\theta_{N,3}^{C}-\widehat\theta_N
=
o_{P_{\theta,\Gamma}}(N^{-3/4}),
\]
together with the same uniform derivative bounds used in Section~\ref{subsubsec:proof-calibrated-rate-leading}, the normalized remainder satisfies
\[
o_{P_{\theta,\Gamma}}(N^{-1}).
\]
Moreover, the first endpoint Hessian inherits the strict curvature of the pooled criterion, while the Hessian at $\widehat\theta_{N,3}^{C}$ is negative semidefinite by the second-order necessary condition for the statistically relevant local maximizer. Hence the normalized average Hessian is uniformly nonsingular with probability tending to one. Multiplying by its uniformly bounded inverse therefore yields
\[
\widehat\theta_{N,3}^{C}
-
\widehat\theta_N
=
o_{P_{\theta,\Gamma}}(N^{-1}).
\]

\section{Full Experimental Results}\label{sec: full experimental results}

In this section, we present more extensive simulation results for the single-round estimator.

\subsection{Bounded vs Unbounded Mini Surrogate Optimization}\label{sec: bounded vs unbounded}

The experiments in Section~\ref{subsec:sim-cox} employed a $[-5,5]$ bound on the regression coefficient of each covariate in the mini-surrogate estimation step, as was done in~\cite{chendistributed}. This was largely driven by the coefficient of the sixth covariate, which took the value 1 with low probability and would thus take values with magnitude over 15 on a not-insignificant number of instances compared to a true value of -0.6. Table~\ref{tab:cox_bounds_comparison} compares the performance of the one-shot method when the mini-surrogate estimation is unbounded. The distances of the one-shot estimator to the pooled estimator, the root mean squared error to the pooled estimator, the mean standard deviation, and the mean distance to the true parameter are shown.

\begin{table}[ht]
\centering
\small
\begin{tabular}{lllcccc}
\hline
Coordination & Estimator & Optimization & $L_2$ Approx. Error & RMSE & SD & $L_2$ Est. Error \\
\hline
\multicolumn{7}{l}{\textbf{Pooled estimator}}\\
 & Pooled & -- & -- & -- & 104.797 & 251.667 \\
\hline
Leading site & Quadratic & Bounded & 345.290 & 385.920 & 121.063 & 406.602 \\
 &  & Unbounded & 356.296 & 401.134 & 122.813 & 416.273 \\
\hline
 & Quadratic + Cal. & Bounded & 185.611 & 236.274 & 117.596 & 311.888 \\
 &  & Unbounded & 223.915 & 291.283 & 123.637 & 340.514 \\
\hline
 & Cubic & Bounded & 193.568 & 237.917 & 121.216 & 316.849 \\
 &  & Unbounded & 234.974 & 295.530 & 127.659 & 347.262 \\
\hline
 & Cubic + Cal. & Bounded & 123.386 & 165.802 & 110.251 & 278.395 \\
 &  & Unbounded & 203.661 & 279.836 & 121.308 & 330.990 \\
\hline
Central server & Quadratic & Bounded & 597.785 & 604.786 & 94.150 & 609.181 \\
 &  & Unbounded & 623.605 & 631.306 & 93.904 & 632.493 \\
\hline
 & Quadratic + Cal. & Bounded & 380.431 & 387.985 & 96.521 & 416.573 \\
 &  & Unbounded & 466.886 & 477.030 & 95.115 & 487.694 \\
\hline
 & Cubic & Bounded & 362.141 & 369.561 & 97.341 & 402.360 \\
 &  & Unbounded & 457.077 & 467.433 & 95.554 & 479.075 \\
\hline
 & Cubic + Cal. & Bounded & 265.616 & 271.034 & 97.829 & 331.295 \\
 &  & Unbounded & 445.140 & 455.862 & 94.290 & 467.551 \\
\hline
\end{tabular}
\caption{Comparison of bounded and unbounded first-stage optimization in the Cox regression experiments. All reported statistics are multiplied by $10^3$.}
\label{tab:cox_bounds_comparison}
\end{table}

As can be seen, using the bounds during the first-stage optimization significantly improved the performance of the method, as the final surrogate functions will be more accurate the closer the expansion points are to each other. The calibrated cubic estimator is still the best performing method when the optimization is unbounded, but the improvement over the uncalibrated cubic estimator is significantly reduced.

\begin{figure}[htbp]
\centering
\includegraphics[width=\textwidth]{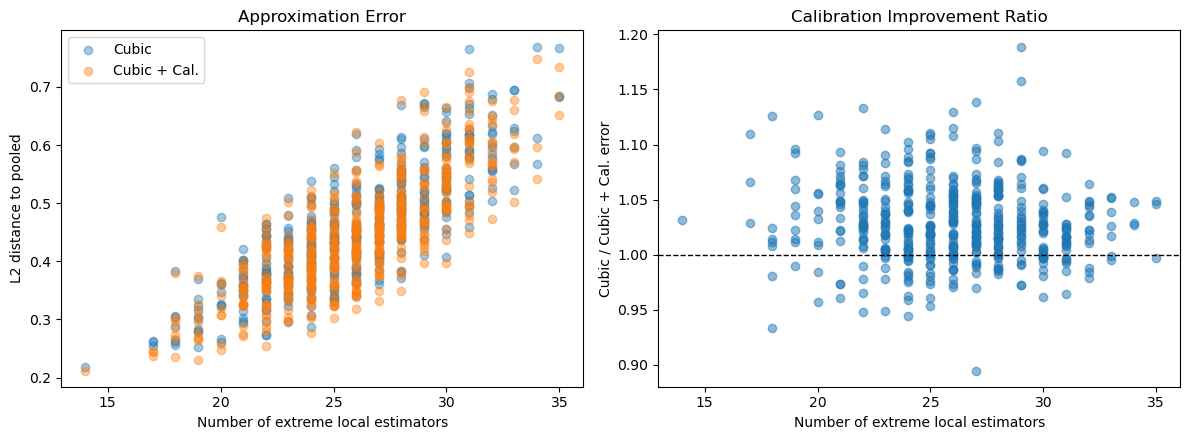}
\caption{Relationship between the number of extreme local estimates and the approximation error in the Cox regression experiments. The left panel shows the $L_2$ distance to the pooled-data estimator for the cubic and calibrated cubic estimators. The right panel shows the ratio of cubic approximation error to calibrated cubic approximation error, with values greater than one indicating improvement from calibration.}
\label{fig:extreme_vs_error}
\end{figure}

To further investigate the relationship between the stability of the mini-surrogate estimators and the accuracy of the resulting estimators, Figure~\ref{fig:extreme_vs_error} plots the approximation error against the number of extreme local estimates in each Monte Carlo replication, where extreme denotes any estimator with at least one coefficient greater than 5 in absolute value. We observe a strong positive association between the number of extreme local estimates and the approximation error for both the cubic and calibrated cubic estimators. For the cubic estimator, the Pearson and Spearman correlations are 0.812 and 0.803, respectively, while the corresponding correlations for the calibrated cubic estimator are 0.799 and 0.786. These results suggest that greater instability among the local expansion points is associated with reduced accuracy of the final estimator, consistent with the sensitivity of higher-order Taylor approximations to the locations of their expansion points and the intrinsic difficulty associated with local sites being more heterogeneous.

The right panel of Figure~\ref{fig:extreme_vs_error} further illustrates the effects of unstable mini-surrogate estimation on the calibration procedure. While the number of extreme local estimates exhibits only a weak negative correlation with the relative improvement from calibration (Pearson $=-0.088$, Spearman $=-0.103$), the improvement ratios are concentrated near one, indicating that calibration often provides only modest additional improvements in these settings. This contrasts with the median improvement ratio of 1.357 observed under bounded optimization. Taken together, these findings suggest that instability in the mini-surrogate estimators can diminish not only the accuracy of the initial cubic approximation, but also the additional benefits obtained through calibration.

\subsection{Second vs Third-Order Mini Surrogate}

The leading-site procedure permits the use of either second- or third-order mini-surrogates to construct the local expansion points. Although the primary experiments employ quadratic mini-surrogates, we additionally investigate whether incorporating third-order information at this preliminary stage improves the accuracy of the resulting estimators. Specifically, we compare quadratic and cubic mini-surrogates across the logistic regression and Cox regression experiments, while keeping the subsequent surrogate construction and calibration procedures unchanged within each experimental setting.

In the bounded Cox experiments, the initial quadratic mini-surrogate is optimized subject to box constraints, whereas the subsequent third-order Newton mini-surrogate optimization is unconstrained. Consequently, the resulting cubic mini-surrogate estimator is not itself restricted to the prescribed bounds. Nevertheless, the bounded and unbounded procedures can produce different cubic mini-surrogate estimates because the constraints affect both the initial quadratic solution and the leading-site estimator used to construct the mini-surrogate. In particular, the leading-site estimator determines the expansion point and the second- and third-order derivative information incorporated into the cubic mini-surrogate. Thus, the bounded and unbounded experiments generally involve different cubic mini-surrogate objectives and initializations, even though the final third-order Newton refinement is unconstrained in both cases.

The resulting comparisons are reported in Table \ref{tab:mini_surrogate_comparison}.

\begin{table}[ht]
\centering
\small
\begin{tabular}{llcccc}
\hline
Estimator & Mini & $L_2$ Approx. & RMSE & SD & $L_2$ Est. \\
\hline
\multicolumn{6}{l}{\textbf{Logistic}}\\
\hline
Quadratic & Quad & 41.515 & 43.932 & 31.774 & 83.200 \\
 & Cubic & 42.598 & 45.757 & 32.010 & 83.956 \\
\hline
Quadratic + Cal. & Quad & 2.568 & 3.566 & 31.477 & 74.859 \\
 & Cubic & 2.772 & 3.985 & 31.479 & 74.857 \\
\hline
Cubic & Quad & 2.552 & 3.623 & 31.489 & 74.892 \\
 & Cubic & 2.661 & 3.781 & 31.486 & 74.877 \\
\hline
Cubic + Cal. & Quad & 0.188 & 0.312 & 31.419 & 74.764 \\
 & Cubic & 0.202 & 0.351 & 31.419 & 74.764 \\
\hline
\multicolumn{6}{l}{\textbf{Cox (bounded)}}\\
\hline
Quadratic & Quad & 345.290 & 385.920 & 121.063 & 406.602 \\
 & Cubic & 361.351 & 402.095 & 122.468 & 418.671 \\
\hline
Quadratic + Cal. & Quad & 185.611 & 236.274 & 117.596 & 311.888 \\
 & Cubic & 198.054 & 253.187 & 118.601 & 319.576 \\
\hline
Cubic & Quad & 193.568 & 237.917 & 121.216 & 316.849 \\
 & Cubic & 201.288 & 251.040 & 120.949 & 321.166 \\
\hline
Cubic + Cal. & Quad & 123.386 & 165.802 & 110.251 & 278.395 \\
 & Cubic & 142.507 & 198.712 & 114.370 & 289.006 \\
\hline
\multicolumn{6}{l}{\textbf{Cox (unbounded)}}\\
\hline
Quadratic & Quad & 356.296 & 401.134 & 122.813 & 416.273 \\
 & Cubic & 362.526 & 404.936 & 122.288 & 419.942 \\
\hline
Quadratic + Cal. & Quad & 223.915 & 291.283 & 123.637 & 340.514 \\
 & Cubic & 223.617 & 290.580 & 122.695 & 339.356 \\
\hline
Cubic & Quad & 234.974 & 295.530 & 127.659 & 347.262 \\
 & Cubic & 229.140 & 290.805 & 125.311 & 342.218 \\
\hline
Cubic + Cal. & Quad & 203.661 & 279.836 & 121.308 & 330.990 \\
 & Cubic & 205.436 & 285.675 & 123.483 & 332.318 \\
\hline
\end{tabular}
\caption{Comparison of quadratic and cubic mini-surrogates under leading-site coordination across logistic regression and bounded and unbounded Cox regression experiments. $L_2$ Approx. denotes mean Euclidean distance to the pooled estimator, RMSE denotes root mean squared distance to pooled, SD denotes mean coefficient-wise standard deviation, and $L_2$ Est. denotes mean Euclidean distance to the true parameter. All statistics are multiplied by $10^3$.}
\label{tab:mini_surrogate_comparison}
\end{table}

Across the three experimental settings, the use of cubic mini-surrogates provides little additional benefit over quadratic mini-surrogates. In the logistic regression experiment, the resulting estimators exhibit nearly identical approximation accuracy, with the quadratic mini-surrogates performing slightly better. The quadratic mini-surrogates also yield consistently smaller approximation errors in the bounded Cox experiment, while the unbounded Cox results are more mixed, with relatively small differences between the two approaches.

These findings suggest that incorporating third-order information into the preliminary mini-surrogate estimation does not necessarily translate into improved accuracy of the final estimator. One possible explanation is that the additional optimization required by the cubic mini-surrogates introduces numerical sensitivity without providing a sufficiently large improvement in the local expansion points. However, the present experiments do not establish this mechanism. Given the comparable performance and lower computational complexity of quadratic mini-surrogates, we adopt them for the primary simulation comparisons.

\end{document}